\documentclass{aa}

\usepackage{graphicx}
\usepackage{booktabs}
\usepackage{placeins}

\usepackage{txfonts}
\usepackage{hyperref}
\hypersetup{
        colorlinks=true,
        breaklinks=true,
        citecolor=blue,
        allcolors=blue
}

\makeatletter
\renewcommand*\aa@pageof{, page \thepage{} of \pageref*{LastPage}}
\makeatother

\newcommand{\karin}[1]{{\color{black} #1}}

\begin{document}

   \title{The dynamical impact of cosmic rays in the Rhea magnetohydrodynamic simulations}

   \author{Karin~Kjellgren
          \inst{1}
          \and
          Philipp~Girichidis\inst{1}
          \and
          Junia~Göller\inst{1}
          \and
          Noé~Brucy\inst{1,2}
          \and 
          Ralf~S. ~Klessen\inst{1,3,4,5}
          \and
          Robin~G.~Tress\inst{6}
          \and
          Juan~D.~Soler\inst{7}
          \and
          Christoph~Pfrommer\inst{8}
          \and
          Maria~Werhahn\inst{9}
          \and
          Simon~C.~O.~Glover\inst{1}
          \and
          Rowan~Smith\inst{10}
          \and
          Leonardo~Testi\inst{11}
          \and
          Sergio~Molinari\inst{7}
          }

   \institute{Universit\"{a}t Heidelberg, Zentrum f\"{u}r Astronomie, Institut f\"{u}r Theoretische Astrophysik, Albert-Ueberle-Str.\ 2, 69120 Heidelberg, Germany
   \and
   Centre de Recherche Astrophysique de Lyon UMR5574, ENS de Lyon, Univ. Lyon1, CNRS, Université de Lyon, 69007, Lyon, France
   \and 
   Universit\"{a}t Heidelberg, Interdisziplin\"{a}res Zentrum f\"{u}r Wissenschaftliches Rechnen, Im Neuenheimer Feld 225, 69120 Heidelberg, Germany
   \and
   Harvard-Smithsonian Center for Astrophysics, 60 Garden Street, Cambridge, MA 02138, U.S.A. \label{CfA}
   \and
   Elizabeth S. and Richard M. Cashin Fellow at the Radcliffe Institute for Advanced Studies at Harvard University, 10 Garden Street, Cambridge, MA 02138, U.S.A. \label{Radcliffe}
   \and
   Institute of Physics, Laboratory for Galaxy Evolution and Spectral Modelling, EPFL, Observatoire de Sauverny, Chemin Pegasi 51, 1290 Versoix, Switzerland
   \and
   Istituto di Astrofisica e Planetologia Spaziali (IAPS), INAF, Via Fosso del Cavaliere 100, 00133 Roma, Italy
   \and
   Leibniz-Institut f\"{u}r Astrophysik Potsdam (AIP), 
   An der Sternwarte 16, D-14482 Potsdam, Germany
   \and
   Max-Planck-Institut f\"{u}r Astrophysik , Karl-Schwarzschild-Str. 1, 85748 Garching, Germany
   \and
   SUPA, School of Physics and Astronomy, University of St Andrews, North Haugh, St Andrews, KY16 9SS, U.K.
   \and
   Dipartimento di Fisica e Astronomia, Universit\`{a} di Bologna, Via Gobetti 93/2, 40122 Bologna, Italy
}

   \date{}
 
  \abstract
   {This study explores the dynamical impact of cosmic rays (CRs) in Milky Way-like galaxies using the Rhea simulation suite. Cosmic rays, with their substantial energy density, influence the interstellar medium (ISM) by supporting galactic winds, modulating star formation, and shaping ISM energetics. The simulations incorporate a multiphase ISM, self-consistent CR transport in the advection-diffusion approximation, and interactions with magnetic fields to study their effects on galaxy evolution.
Key findings reveal that CRs reduce star formation rates (SFRs) and drive weak, but sustained outflows with mass-loading factors of $\sim0.2$, transporting a substantial fraction (20\%-60\%) of the injected CR energy. These CR-driven outflows are launched not just from the galactic center, but across the entire disk, illustrating their pervasive dynamical influence. Galactic disks supported by CRs exhibit broader vertical structures compared to magnetic-field-dominated setups, although the scale heights are similar. CR feedback enhances magnetic flux transport to the circumgalactic medium (CGM), leading to a magnetically enriched CGM with field strengths of $\sim0.5\mu\mathrm{G,}$ while reducing gas temperatures to $\lesssim10^5\,\mathrm{K}$. The CR energy is relatively smoothly distributed in the disk, with gradient lengths exceeding the typical size of molecular clouds, indicating that the CR behavior is not adiabatic.}

   \keywords{Cosmic rays --- Magnetohydrodynamics (MHD) --- ISM: jets and outflows --- Galaxies: evolution --- Galaxies: magnetic fields}

   \maketitle
%
%-------------------------------------------------------------------
%\nocite{tange_ole_2018_1146014}
\section{Introduction}
Cosmic rays (CRs) are charged particles, which  are predominantly accelerated via diffusive shock acceleration \citep[DSA;][]{Krymskii1977, AxfordEtAl1978, Bell1978a, Bell1978b, BlandfordOstriker1978}. In the interstellar medium (ISM), the main accelerators are supernova remnants (SNRs), in which CRs reach relativistic speeds. CRs are an integral component of the ISM and might hold the key for some of the unanswered questions in galaxy formation and evolution. The energy density of CRs is roughly in equipartition with the thermal, magnetic, and kinetic components \citep{Ferriere2001,cox2005,naab&ostriker2017}, meaning they can have a significant dynamical impact on the ISM \citep{Strong2007,grenier2015,Klessen2016}. They heat and ionize gas, thereby regulating star formation \citep[e.g.,]{Uhlig2012,Butsky_2018,Dashyan_2020},  playing a central role in astrochemistry \citep{Albertsson2018,Padovani2018,Phan2018,PadovaniEtAl2020}. They have also been suggested to be able to launch and support galactic outflows \citep{Breitschwerdt1991,dorfi_breitschwerdt2012, Booth2013, Salem2014,Girichidis2016, girichidis_2018,Ruszkowski2017}, which are ubiquitous in galaxies both in our nearby Universe and at higher redshift as well \citep{Veilleux_2005,steidel2010}. 

Feedback processes in galaxies heat and eject gas, thereby lowering the amount of available cold gas to form stars. Effective stellar feedback is needed to explain the low star formation efficiencies that we observe in typical spiral galaxies \citep{Moster2013,krumholz_2014, utomo_2018}. Thermally driven feedback, such as energy input from stellar winds or supernovae (SNe), has the problem of efficient cooling, which lessens its impact, particularly in environments with high gas densities. CRs on the other hand have much longer cooling times compared to radiative cooling of thermal plasma, and are dynamically coupled to the gas through their scattering off of gyroresonant electromagnetic waves, allowing them to impart energy and momentum to the ISM (see, e.g., \citealt{ruszkowski2023} for a review). This, together with the rapid diffusion of CRs out of the disk, allows them to successfully transport SN energy to regions of low-density gas above the disk, where they can establish pressure gradients that can accelerate and sustain galactic winds.

The dynamical impact of CRs and, in particular, their ability to launch outflows has been demonstrated in numerical simulations using a variety of setups. Simulations on interstellar scales that have the ability to resolve the multiphase ISM \citep{Girichidis2016,girichidis_2018, Simpson2016,Simpson2023,rathjen2021,Thomas2024,Sike2024}, models of isolated galaxies \citep{Uhlig2012,Hanasz2013,pakmor2016,Jacob_2017,Butsky_2018,Dashyan_2020,Girichidis2022,peschken2021, Girichidis2024,Thomas2023}, and cosmological zoom-in simulations \citep{Salem2014, Hopkins2020, Buck_2020, montero2023, DeFelippis_2024} have all reported that CRs thicken the galactic disk, reduce star formation, and support galactic winds.

As CRs gyrate and move along magnetic field lines, they transport the CRs relative to the thermal gas, allowing them to launch outflows. The specifics of the transport of CRs have a big impact on the efficacy of  CR feedback \citep{Wiener2017,Butsky_2018,Dashyan_2020,Semenov2021}. However, the physical details of this process are not yet fully understood. A gray or ``one-moment'' approach is often adopted, where only the total integrated CR energy is evolved and the CR population is fully described by only the CR energy \citep{HanaszStrongGirichidis2021, ruszkowski2023}. The CRs are then transported via diffusion or streaming, where a constant diffusion coefficient is assumed in the case of the former. However, this scenario fails to account for all the complexities of CR transport, which might be important. For example, the diffusion coefficient, which determines the rate of CR transport, could vary in different phases of the ISM \citep{Armillotta_2021,armillotta2024}. For these reasons, there have been efforts to implement a more realistic two-moment approach, where the diffusion coefficient is self-consistently calculated and  CR transport allows for a combination of streaming and diffusion \citep{Jiang2018,thomas2019,Thomas2023}. There are now also numerical simulations that move away from the gray approach by evolving the full CR spectra to account for the energy-dependence of CR transport \citep{girichidis_I_2020, Girichidis2022, Girichidis2024,Hopkins2022}. 

\karin{Observational constraints provide important insights into CR transport and feedback processes, which informs and complements numerical simulations.} Galactic CRs can be directly detected at Earth, which allows  for their transport processes to be probed \citep{Strong2007,Evoli2008,Kissmann2014,Werhahn_2021_I,Hopkins2022}. In addition, they can be indirectly detected through their interaction with the ambient gas. Neutral pions are created in collisions between CR protons and gas particles, which quickly decay into gamma rays. In particular, the far-infrared (which probes the star formation) and gamma-emission relation allows for a probing of the dynamical impact of CRs through the calorimetric fraction, which is the fraction of CR energy that is lost to emission before the CRs have time to escape their host galaxies. This relation has also been successfully modeled in simulations \citep{Pfrommer2017_gamma,Chan_2019, Buck_2020, Werhahn_2021_II,Werhahn2023}\karin{,  demonstrating how numerical approaches can help us interpret gamma ray observations to understand CR physics \citep[however, see][for challenges in interpreting local CR data]{Kempski2022}}. Thanks to instruments such as Fermi LAT, we have access to numerous observations of the diffuse gamma emission in our own Galaxy and from nearby galaxies. In contrast to the gamma-ray observations, radio telescopes are much more sensitive and enable observations at greater angular resolution. This makes it a prime observational tool for inferring CR transport properties via scaling relations and the individual spectra of radio intensity \citep{Thompson2006,Lacki2010,Werhahn_2021_III,Pfrommer2022} as well as polarization \citep{Chiu2024}. However, the radio synchrotron emission is degenerate with the magnetic field strength. It remains challenging to separate the thermal free-free emission and absorption from the intrinsic non-thermal radio emission, which probes CR physics. 

The wealth of observations available for comparison to makes the Milky Way an especially interesting object of study. This was the motivation behind the Rhea simulations, a simulation suite of isolated, Milky Way-like galaxies that has been carefully designed to reproduce key qualities of the Milky Way. The galaxies are all main sequence spiral galaxies of similar mass and size to the Milky Way, with star formation rates (SFRs) that match what is observed in the Milky Way \citep{göller2024}. The simulations include a multiphase ISM that allows us to study how the dynamics are affected at different locations in the disk (e.g.,\ as introduced by \citealt{Tress2020} or \citealt{Sormani2020}). In this paper,  the first in a series, we analyze four simulations from this suite to specifically investigate the dynamical impact of CRs on Milky-Way-like galaxies. 

The outline of the paper is as follows. In Section \ref{sec:method}, we describe the setup of our simulations of Milky Way-like galaxies and the implementation of CRs. In Section \ref{sec:disk prop}, we start presenting our results by investigating the impact of CRs on global disk properties such as morphology, ISM structure, and star formation. We discuss the different energy components in the ISM and circumgalactic medium (CGM) in Section \ref{sec:energetics}. 
The outflows in our galaxies and their impact on vertical structure are analyzed in Section \ref{sec:outflows}, where we also investigate important properties such as the mass and energy loading factors of the winds. We present  a discussion in Section \ref{sec:discussion} and our conclusions in Section \ref{sec:conclusions}. 

\section{Simulation setup}\label{sec:method}

\subsection{The Rhea simulations}
The simulations analyzed in this work are a part of the Rhea suite of simulations of Milky Way-like galaxies. Full details of the Rhea simulations are provided by \citet{göller2024}. Here, we only summarize the most important points of the setup, the chemistry, and the implementation of star formation and stellar feedback in the form of SNe. 

All Rhea simulations were performed using \textsc{Arepo} \citep{arepo,Pakmor2016arepo,arepo_public}, a magnetohydrodynamical (MHD) code implemented on a moving mesh. \textsc{Arepo} assumes ideal MHD in a non-cosmological environment, which has been found to result in magnetic fields in Milky Way-sized galaxies comparable to observational inferences \citep[e.g.,][]{pakmor2017,Pakmor2020}. To ensure stability in more dynamic environments, the code employs a Powell eight-wave scheme for divergence cleaning \citep{Powell1999, Pakmor2011,Pakmor2013}.
Together with a two-fluid approximation composed of thermal gas and CRs, the fundamental conservation laws that \textsc{Arepo} solves can be written in Gaussian cgs units as \citep{Pfrommer2017}:

\begin{align}
&\frac{\partial \rho}{\partial t}+\boldsymbol\nabla\boldsymbol\cdot(\rho\boldsymbol{v})=0,\\
&\frac{\partial(\rho\boldsymbol{v})}{\partial t}+\boldsymbol\nabla\boldsymbol\cdot\left(\rho\boldsymbol{v}\boldsymbol{v}^\mathrm{T}+P\mathbf{I}-\frac{\boldsymbol{B}\boldsymbol{B}^\mathrm{T}}{4\pi}\right)=-\rho\boldsymbol{\nabla}\Phi,\\
&\begin{aligned}
    &\frac{\partial e}{\partial t}+\boldsymbol\nabla\boldsymbol\cdot\left[(e+P)\boldsymbol{v}-\frac{\boldsymbol{B}(\boldsymbol{v}\boldsymbol\cdot\boldsymbol{B})}{4\pi}\right]=\\
    &-\rho(\boldsymbol{v}\boldsymbol{\cdot}\boldsymbol{\nabla}\Phi)+ P_{\mathrm{cr}}\boldsymbol\nabla\boldsymbol\cdot\boldsymbol{v}-\boldsymbol{v}_{\mathrm{st}}\boldsymbol\cdot\boldsymbol\nabla P_{\mathrm{cr}}+\Lambda_{\mathrm{th}}+\Gamma_{\mathrm{th}},
\end{aligned}\\
&\begin{aligned}
    &\frac{\partial e_{\mathrm{cr}}}{\partial t}+\boldsymbol\nabla\boldsymbol\cdot\left[e_{\mathrm{cr}}\boldsymbol{v}+(e_\mathrm{cr}+P_{\mathrm{cr}})\boldsymbol{v}_{\mathrm{st}}-\kappa\boldsymbol{b}(\boldsymbol{b}\boldsymbol\cdot\boldsymbol\nabla e_{\mathrm{cr}})\right]=\\\label{eq_ecr}
    &-P_{\mathrm{cr}}\boldsymbol\nabla\boldsymbol\cdot\boldsymbol{v}+\boldsymbol{v}_{\mathrm{st}}\boldsymbol\cdot\boldsymbol\nabla P_{\mathrm{cr}}+\Lambda_{\mathrm{cr}}+\Gamma_{\mathrm{cr}},
\end{aligned}\\
&\frac{\partial \boldsymbol{B}}{\partial t}+\boldsymbol\nabla\boldsymbol\cdot(\boldsymbol{B}\boldsymbol{v}^\mathrm{T}-\boldsymbol{v}\boldsymbol{B}^\mathrm{T})=0,
\end{align}
where $\rho$ is the gas density, $\boldsymbol{v}$ is the gas velocity, and $\boldsymbol{B}$ is the magnetic field vector. The unit field vector along the magnetic field is defined as $\boldsymbol{b}=\boldsymbol{B}/|\boldsymbol{B}|$. The gravitational potential $\Phi$ consists of three components summed together: an external flat potential, the self-gravity of the gas, and the potential generated by the star particles, which are described in detail below. The CR pressure and energy density are denoted as $P_{\mathrm{cr}}$ and $e_{\mathrm{cr}}$, respectively, and $\kappa$ signifies the CR diffusion coefficient along the magnetic field. The streaming velocity, $\boldsymbol{v}_{\mathrm{st}}$, is defined as 
\begin{align}
    \boldsymbol{v}_{\rm st} = -\boldsymbol{v}_A \ \mathrm{sign} (\boldsymbol{B}\boldsymbol\cdot\boldsymbol{\nabla}P_{\rm cr})=-\frac{\boldsymbol{B}}{\sqrt{4\pi\rho}}\frac{\boldsymbol{B}\boldsymbol\cdot\boldsymbol{\nabla}P_{\rm cr}}{|\boldsymbol{B}\boldsymbol\cdot\boldsymbol{\nabla}P_{\rm cr}|},
\end{align}
where $\boldsymbol{v}_{\rm A}$ is the Alfvén velocity. Sources and sinks of thermal and CR energy density are denoted by $\Gamma_{\mathrm{th/cr}}$ and $\Lambda_{\mathrm{th/cr}}$, respectively. The total pressure, $P$, and the total energy density, $e$, (excluding CRs) are defined as:
\begin{align}
    P &= P_{\mathrm{th}}+P_\mathrm{cr}+\frac{\boldsymbol{B}^2}{8\pi}\;,\\
    e &= e_\mathrm{th}+\frac{\rho\boldsymbol{v}^2}{2}+\frac{\boldsymbol{B}^2}{8\pi}\;.
\end{align}
The CR energy density, $e_{\rm cr}$, is evolved separately following Equation \ref{eq_ecr}, but without explicitly accounting for CR streaming. However, we account for the effective energy losses due to CR streaming, as explained below. To close the system of equations we needed an equation of state, where we adopted an adiabatic index of $\gamma_{\mathrm{th}}=5/3$ for the thermal component and $\gamma_{\mathrm{cr}}=4/3$ for the relativistic CR component.

The chemistry was modeled using the NL97 non-equilibrium chemical network from \cite{Glover2012}, which follows the hydrogen
chemistry of the ISM and includes a highly simplified treatment of CO formation and destruction based on \citet{Nelson1997}. This network allows us to track the abundances of ionized, atomic, and molecular hydrogen, as well as carbon monoxide and singly ionized carbon, which we used as input for the atomic and molecular cooling function described in \citet{Clark2019}. We adopted a fixed CR ionization rate of $\zeta_{\rm CR}$\,$=$\,$3\times 10^{-17} \: {\rm s^{-1}}$ for atomic hydrogen; values for other species (e.g.,\ H$_{2}$) were derived from this rate by applying appropriate scaling factors \karin{from the UMIST database \citep{leteuff2000}}. We note that the variations in the CR energy density could also alter the CR ionization rate at low CR energies. However, the spectral connection between the GeV CRs that we follow in the simulations and the low-energy CRs responsible for the ionization rate is non-trivial. A simple rescaling based on the energy has been tested in \citet{girichidis_2018} without significant impact on the formation of cold gas. A scaling of the ionization rate based on the total integrated CR energy density in combination with a local column density reflecting the scaling in observations \citep[e.g.,][]{PadovaniEtAl2022} will be left to a future work.

The star formation recipe used in the Rhea simulations employs collisionless star particles. At the resolution used in this study, each of these star particles actually represents a small stellar population. A cell in the simulation is flagged as potentially star-forming once the mass of the cell exceeds one-eighth of its Jeans mass. A star particle is created either if the cell's gas mass exceeds its Jeans mass, or stochastically, with a 
probability, $\mathcal{P,}$ derived as in \cite{springel-hernquist2003}: 
\begin{align}
        \lambda&=\mathrm{SFR}\frac{\mathrm{\Delta t}}{M_{\rm cell}},\\
        \mathcal{P}&=\frac{M_\mathrm{\rm cell}}{M_\mathrm{starP}}\left(1-\exp(-\lambda)\right),
\end{align}
where $\rm \Delta t$ is the timestep of the cell, $M_\mathrm{cell}$ is the mass of gas in the cell, and $M_\mathrm{starP}$ is the desired mass of the star particle, which equals $M_\mathrm{cell}$ in most cases. However, if the mass of the cell exceeds the imposed mass resolution of 3000 $\rm M_\odot$ by more than a factor of 2, so that it would have been split into 2 in the next time step, only half of the cell's mass is converted into stars while the other half remains as gas. The time step of star-forming cells is adjusted so that $\mathcal{P}<1$ always. If a gas cell is converted into a star particle, it is populated with individual stars by sampling from the high-mass end of the Kroupa initial mass function (IMF) using the algorithm described in \citet{Sormani2017}. Stars with masses between $8 \ \rm M_\odot$ and $120 \ \rm M_\odot$ eventually explode as SNe, injecting momentum or energy into their surroundings depending on whether or not the radius of the SNR at the end of the Sedov-Taylor phase ($r_{\rm ST}$) exceeds the injection radius ($r_{\rm inj}$). The injection radius (i.e., the radius of the spherical volume into which the SN feedback is injected) was set to 100 pc, where we ensured a sufficient number of cells within that region. If $r_{\rm ST}>r_{\rm inj}$, an energy of $10^{51}$ erg was injected isotropically into this volume and the chemical composition is immediately changed to fully ionized. Otherwise, no thermal energy is injected; instead, a momentum of $p=2.6\times10^5\left(\frac{n}{\rm cm^{-3}}\right)^{-2/17} \rm \ M_\odot \ km \ s^{-1}$ (derived in e.g.,\ \citealt{Gatto_2015}) was injected into the cells in the injection region, where $n$ is the gas number density.
\begin{table}[t]
    \centering
    \begin{tabular}{lll}
    \toprule
     Name & CRs & $B_0$ \\
     & &  [$\mu$G]\\
     \midrule
     MHD & no & 3 \\
     MHD-low & no &  $3\times10^{-3}$ \\
     CRMHD & yes &  3 \\
     CRMHD-low & yes & $3\times10^{-3}$\\
     \bottomrule \\
    \end{tabular}
    \caption{Properties of simulations analyzed in this work.}
    \label{tab:simulations}
\end{table}

\subsection{Cosmic rays}
The main difference between the simulations analyzed in this work and those in \citet{göller2024} is that our new simulations include CRs and magnetic fields. In this section, we  describe the implementation of CRs in our simulations, largely following the procedure proposed by \cite{Pfrommer2017}. As previously mentioned, CRs are included as a second fluid in addition to the thermal gas, where the CR fluid is modeled with a constant adiabatic index of $\gamma_{\mathrm{cr}}=4/3$. This is a so-called gray approach, where we only followed the momentum-integrated total CR (proton) energy density. \karin{Due to the computational cost of a CR transport model that takes the energy-dependence into account, we  leave that to a future work and discuss more sophisticated approaches in Section~\ref{subsec: caveats}.}

The term $\Gamma_{\rm cr}$ in Equation \eqref{eq_ecr} represents sources of CRs, which we took to be the CR injection in SNe to account for unresolved subgrid DSA at SNR shocks. In each SN, we injected $10^{50}$ erg as CRs (i.e.,\ 10\% of the total explosion energy) into the same cells into which we inject the thermal energy.

We accounted for both the advection and diffusion of CRs, where diffusion occurs parallel to the magnetic field using a constant diffusion coefficient of $\kappa=4\times10^{28}$ $\mathrm{cm}^2 \mathrm{s}^{-1}$. This value is consistent with what is found from comparing to observational data in the Milky Way, which leads to an estimated diffusion coefficient of $\rm (3-5) \times10^{28} \ cm^2s^{-1}$ for CRs at $\sim1$ GeV \citep{Strong2007}. We note, however, that this number depends on several assumptions and might be highly degenerate.

The term $\Lambda_{\rm cr}$ represents the energy sinks of CRs, where we account for hadronic and Coulomb losses. The most important hadronic interaction is a CR proton interacting with a thermal proton (either single or bound in a heavier nucleon) and generating pions, which decay and produce various secondaries as well as gamma rays. The hadronic loss rate assumes a uniform CR spectrum where injection balances the various losses, yielding:
\begin{equation}
    \Lambda_{\mathrm{hadr}}=-7.44\cdot10^{-16}\left(\frac{n_{\rm e}}{\mathrm{cm}^{-3}}\right)\left(\frac{e_{\mathrm{cr}}}{\mathrm{erg \ cm^{-3}}}\right)\mathrm{erg} \ \mathrm{s}^{-1} \ \mathrm{cm}^{-3},
\end{equation}
where $n_{\rm e}$ is the number density of free electrons. Unlike Coulomb losses, most of the energy lost to hadronic interactions escapes as gamma rays and neutrinos instead of heating the gas. Specifically, $5/6$ of the energy is radiated, while the rest goes into heating. A CR ion that is deflected by the Coulomb field of an electron in the background plasma will be accelerated, resulting in bremsstrahlung emission, which cools the ion. In \textsc{Arepo}, the Coulomb loss rate of a CR population is defined as \citep{Pfrommer2017}
\begin{equation}
    \Lambda_{\mathrm{Coul}}=-2.78\cdot10^{-16}\left(\frac{n_{\rm e}}{\mathrm{cm}^{-3}}\right)\left(\frac{e_{\mathrm{cr}}}{\mathrm{erg \ cm^{-3}}}\right)\mathrm{erg} \ \mathrm{s}^{-1} \ \mathrm{cm}^{-3}\;.
\end{equation}
All of the energy lost in Coulomb interactions goes into heating the surrounding ISM.

In addition, we also accounted for Alfvén cooling, which is a way of emulating the losses due to streaming, which are not explicitly included \citep{Wiener2017}. Diffusion is an energy-conserving process, while streaming is not. The CR streaming instability resonantly excites Alfvén and whistler waves \citep{Kulsrud1969,Shalaby_2021,Shalaby_2023}, which scatter the CRs in pitch angle. In a steady state, these waves are damped by some process (such as ion-neutral damping or non-linear Landau damping), which will transfer energy to the gas \citep[see][for an extended review of the underlying physics]{ruszkowski2023}. Therefore, the streaming instability is an indirect way for CRs to transfer their energy to the gas. Streaming effectively reduces the mean CR transport speed $\boldsymbol{v}_{\rm st}$ to the Alfvén speed $\boldsymbol{v}_\mathrm{A}$. The cooling rate is expressed as \citep{wiener2013}
\begin{equation}
    \Lambda_\mathrm{A}=\boldsymbol{v}_\mathrm{st}\boldsymbol\cdot \boldsymbol\nabla P_{\mathrm{cr}}\;.
\end{equation}

\subsection{Simulation details}\label{subsec:sim details}
For each simulated galaxy, we start with a smooth gaseous disk and no stars, with a density distribution given by
\begin{align}
    \rho_\mathrm{gal} \left(R_\mathrm{gal}, z\right)=\frac{\Sigma_0}{4z_{\rm d}}\mathrm{exp}\left(-\frac{R_{\rm m}}{R_\mathrm{gal}}-\frac{R_\mathrm{gal}}{R_{\rm d}}\right)\mathrm{sech}^2\left(\frac{z}{2z_{\rm d}} \right),
    \label{equ:density_profile}
\end{align}
in cylindrical coordinates \citep{McMillan2017, Sormani2019}, with $z_{\rm d}=85\,$pc, $R_{\rm d}=7\,$kpc, $R_{\rm m}=1.5\,$kpc, and $\Sigma_0=50\,\mathrm{M}_\odot/\mathrm{pc}^2$. The total gas mass is $1.2\times10^{10}\ \rm M_\odot$, making all our galaxies Milky Way-like \citep[e.g.,][]{BlandHawthorn2016}. The velocity of the cells is imposed by an external flat gravitational potential that does not include spiral arm or bar features and which results in a flat velocity curve \citep{binney2008galactic}. Our simulation box has a volume of $150^3 \ \rm kpc^3$ and periodic boundary conditions. Initially, all the gas in the box has a temperature of $10^4$ K, \karin{with 99.9\% of the gas mass being located in the disk ($r\leq30$ kpc, $h\leq1$ kpc)}. The mass resolution of our simulations is 3000 $\rm M_\odot$ and the minimum and maximum allowed cell volume is 1 $\rm pc^3$ and 2 $\rm kpc^3$, respectively; except in the disk ($r\leq30$ kpc, $h\leq1$ kpc), where the maximum cell volume is limited to $\rm (100 \  \mathrm{pc})^3$. Additional volume refinement must be applied if the maximum cell volume is exceeded, as described in more detail in Appendix~\ref{app:resolution}. 

All of our simulations include magnetic fields. The magnetic field was initialized as a purely toroidal field, with an initial field strength, $B_0$, of either $3 \ \mu \rm G$ or $3 \  \rm nG$ at a density of $\rho_0 = 10^{-24} \ \rm g \ cm^{-3}$ (i.e., at the mean ISM density). We want to investigate the impact that the magnetic field has on the CR transport and, for that reason, we considered two different values. In our naming convention, we added "-low" to the names of the simulations that use the weaker (3~nG) magnetic field. The field is scaled with the gas density as $B=B_0(\rho/\rho_0)^\alpha$. For super-Alfv\'{e}nic turbulence, a scaling with $\alpha=2/3$ \citep{Mestel1966} is expected. Sub-Alfv\'{e}nic motions result in a scaling with $\alpha=1/2$ \citep{ChandrasekharFermi1953} or $\alpha\le1/2$ \citep{MouschoviasCiolek1999}. Our setup is missing the magnetic field amplification via a turbulent magnetic dynamo driven by the interaction of gas accretion and galactic outflows in the CGM over cosmological times. To avoid  finding artificially strong outflows as a result of the missing magnetic energy in the CGM, we applied a scaling of $\alpha = 1/3$. For the strong field with $B_0=3\,\mu\mathrm{G}$, this results in magnetic field strengths of the order of $1\,\mu\mathrm{G}$ at a height of $10\,\mathrm{kpc}$ above the midplane, which is comparable to cosmological simulations of Milky-Way systems \citep[e.g.,][]{Pakmor2020,PakmorEtAl2024}. We note that the field in this case is not self-consistently amplified via the hydrodynamical evolution. On the contrary, the model with $B_0=3\,\mathrm{nG}$ has a significantly weaker field in the CGM. The resulting outflows can therefore penetrate easier to large heights above the disk. However, the evolution of the simulations, the resulting field strengths and configuration will evolve self-consistently. We further discuss the field strength in Appendix~\ref{app:Bfield_scaling}.

The simulations run for 2 Gyr. During the first gigayear of evolution, the lifetime of stars is increased from a tenth of their normal value to their normal value, after which the simulations run for another gigayear. This is so that more turbulence is induced in the disk early on via SN explosions, which prevents the initially smooth gas disk from immediately collapsing and forming an unphysically large number of stars that end up completely rupturing the galaxy by feedback. Additionally, during the entire evolution, mass return from the SNe is activated. This means that, for each SN explosion, a mass of $M_{\rm starP}/N_{\rm SN}$ is distributed evenly to the cells within the injection radius, where $M_{\rm starP}$ is the mass of the star particle and $N_{\rm SN}$ is the total number of SNe going off in that star particle. This has the consequence that no mass is locked up in the star particles; namely, we do not have gas depletion, nor do we have an old population of star particles. However, we do account for the gravitational effect of an old population of stars via the external potential. This phase is denoted as "phase I" in \citet{göller2024}. Due to the significantly higher computational cost of simulations that include CRs, we ran them for a shorter duration than the simulations analyzed in \citet{göller2024}, focusing only on the first two gigayears of evolution.

This work considers four different simulations. We distinguish between simulations that include CRs (CRMHD) and control runs that do not (MHD). We further differentiate between simulations initialized with a strong and weak magnetic field, where those with a weak field get ``-low'' added to their names. Details of the simulations treated in this work can be found in Table \ref{tab:simulations}.

\section{Galactic disk properties}\label{sec:disk prop}

\begin{figure*}
    \includegraphics[width=\textwidth]{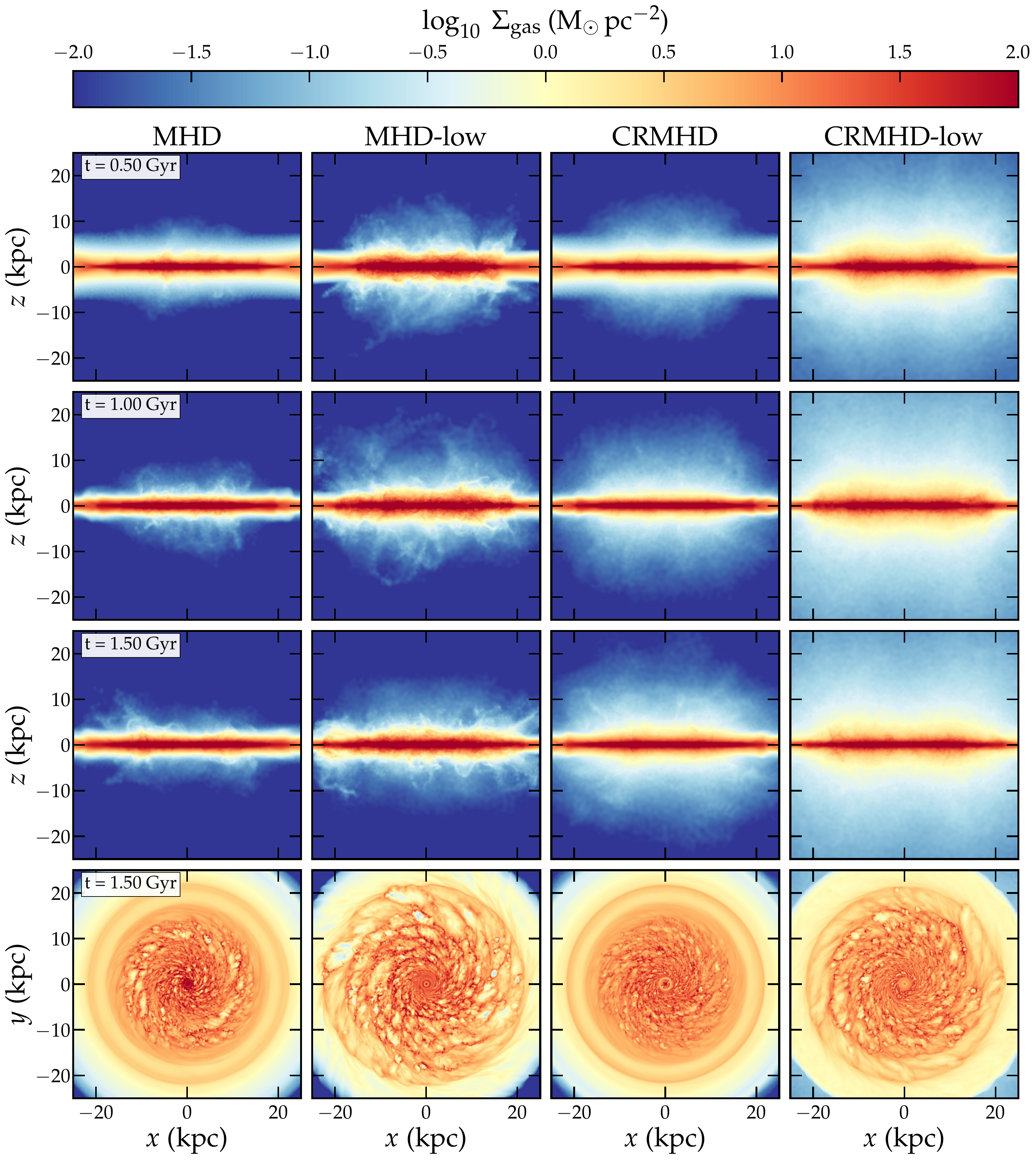}
    \caption{Evolution of our simulations. In each column we show \karin{edge}-on maps of the column density for one simulation at three different times: $t=0.5,1.0$, and $1.5$ Gyr. In the last row, we also show the \karin{face}-on view of the column density at $t=1.5$ Gyr. The simulations with an initially smaller magnetic field develop much larger star-forming disks due to the weaker magnetic support. The inclusion of CRs leads to a more fluffed up disk as the CRs diffuse out and drives a galactic outflow that is stronger in the initially weak-field model CRMHD-low.}
    \label{fig:morphology}
\end{figure*}

\begin{figure*}
    \includegraphics[width=\textwidth]{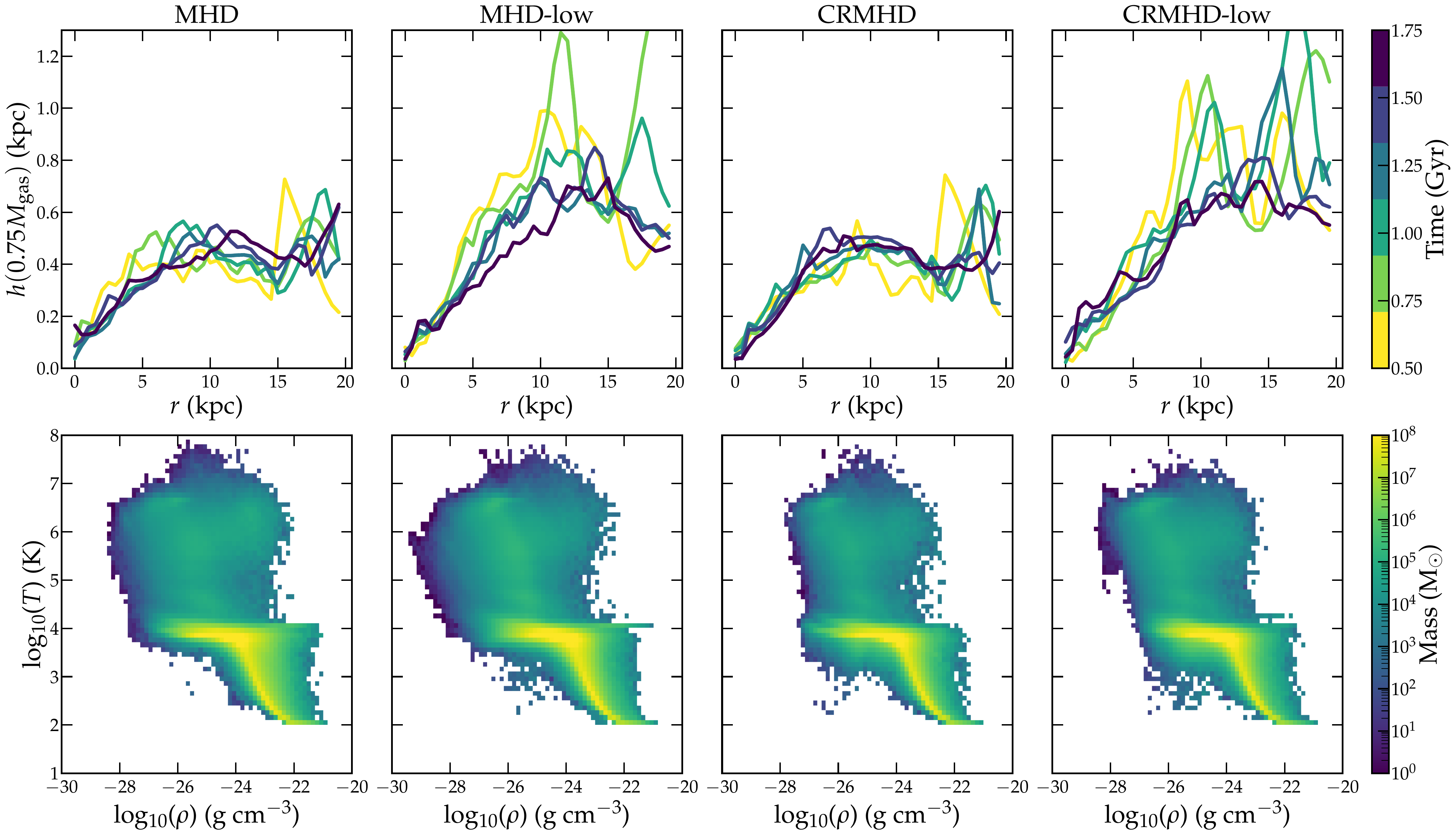}
    \caption{Top: Evolution of the scale height (height above the disk that encloses 75\% of the total gas mass at a given radius) for our four simulations. In all cases the scale height eventually converges, at which point the gas mass is well contained within 1 kpc of the midplane. Bottom: Temperature-density histograms for our simulations at $t=1.5$ Gyr, weighted by the gas mass and limited to a disk of radius of $r=20$ kpc and height of $h=1$ kpc.}
    \label{fig:scaleheight_phaseplot}
\end{figure*}

\subsection{Morphological evolution}
In this section, we offer an overview of the morphological evolution of our model galaxies. In Figure \ref{fig:morphology}, we show projections of the gas density for all our simulations at different points in the evolution. The first three rows depict \karin{edge}-on projections of the galactic disk at $t=0.5,1.0,$ and 1.5 Gyr, respectively. Additionally, in the last row we visualize \karin{face}-on density projections of the galaxies at $t=1.5$ Gyr. We used $t=1.5$ Gyr as our fiducial time for many of the figures presented in this work, as we found that beyond this point there is very little evolution in disk properties.

In \karin{the face-on view of the} galaxies, we observe overdense regions as well as low-density bubbles where SNe have recently exploded. The strength of the magnetic field has a clear impact on the size of the star-forming disk. The CRMHD and MHD galaxies develop star-forming disks of around 15 kpc in radius, whereas the CRMHD-low and MHD-low galaxies develop larger disks with radii closer to 20 kpc. The initially weaker magnetic field in the "-low" simulations results in less magnetic support, which allows for star formation further out from the center. The face-on evolution of the galaxies is marginally affected by the presence of CRs, though there are minor differences. The CRMHD and CRMHD-low galaxies appear somewhat smoother than their MHD-counterparts, with fewer low density regions and overall smaller bubbles. This can be attributed to the CRs smoothing out overdensities in the gas as they diffuse through the disk. However, since our diffusion coefficient is relatively large and the CRs diffuse rapidly this effect is not prominent. Also, the linear diffusion model does not account for ion-neutral damping, which reduces the effect of smoothing \citep{Thomas2024,Sike2024}.

The difference between the simulations, as well as the impact of CRs, is \karin{evident} as we look at the edge-on \karin{evolution of the galaxies}. \karin{Without the inclusion of CR effects, we would develop quite compact disks, with the lower magnetic field in MHD-low allowing for a slightly fluffier disk at $t=1.5$ Gyr. In the simulations with CRs, however, gas is more effectively pushed into the CGM as the CRs diffuse outward and establish gas-lifting pressure gradients above the disk. In the strong $B$-field case, this process is gradual, as seen in the evolution of the edge-on view from $t=0.5-1.5$ Gyr, where gas is slowly lifted to greater heights. In contrast, with the weaker magnetic support in CRMHD-low, the CRs are able to diffuse out the galaxy much earlier.  By $t=0.5$ Gyr, gas has already been lifted further than 20 kpc from the midplane.}

We attempted to quantify this effect by looking at the time evolution of the gas scale height, which we define as the height in a given radial bin that contains 75\% of the total gas mass in that bin. This quantity is illustrated for various times in the top panels of Figure \ref{fig:scaleheight_phaseplot} for all simulations. The CRMHD-low and MHD-low galaxies experience some initial fluctuations, but converge after around 1 Gyr of evolution. These fluctuations are due to the pressure waves previously mentioned, these are also responsible for the peak at around $r=19$ kpc in the CRMHD and MHD simulations. In all four galaxies, most of the gas is well contained within 1 kpc of the midplane, with no substantial differences between simulations with and without CRs, despite the noticeable effects seen in Figure \ref{fig:morphology}. This is because the gas pushed into the CGM is of very low density, therefore, it does not significantly affect the scale height. 

In particular, the MHD and CRMHD galaxies display quite prominent ring-like structures surrounding the star-forming disk, which are magnetosonic waves driven by pressure perturbations in the inner disk as a result of star formation. We consider these features in our analysis and note that their presence does not influence our main conclusions. 

Based on the sizes of the star-forming disks and the scale heights of our galaxies discussed in this section, we define the region of the disk that we discuss later in this work as a cylinder with a radius of 20 kpc and a height of 1 kpc centered on the galactic center; thus,  when we refer to, for instance, averages in the galactic disk, we are referring to averages taken in this region.

\subsection{ISM composition}

\begin{figure*}
    \centering
    \includegraphics[width=0.7\textwidth]{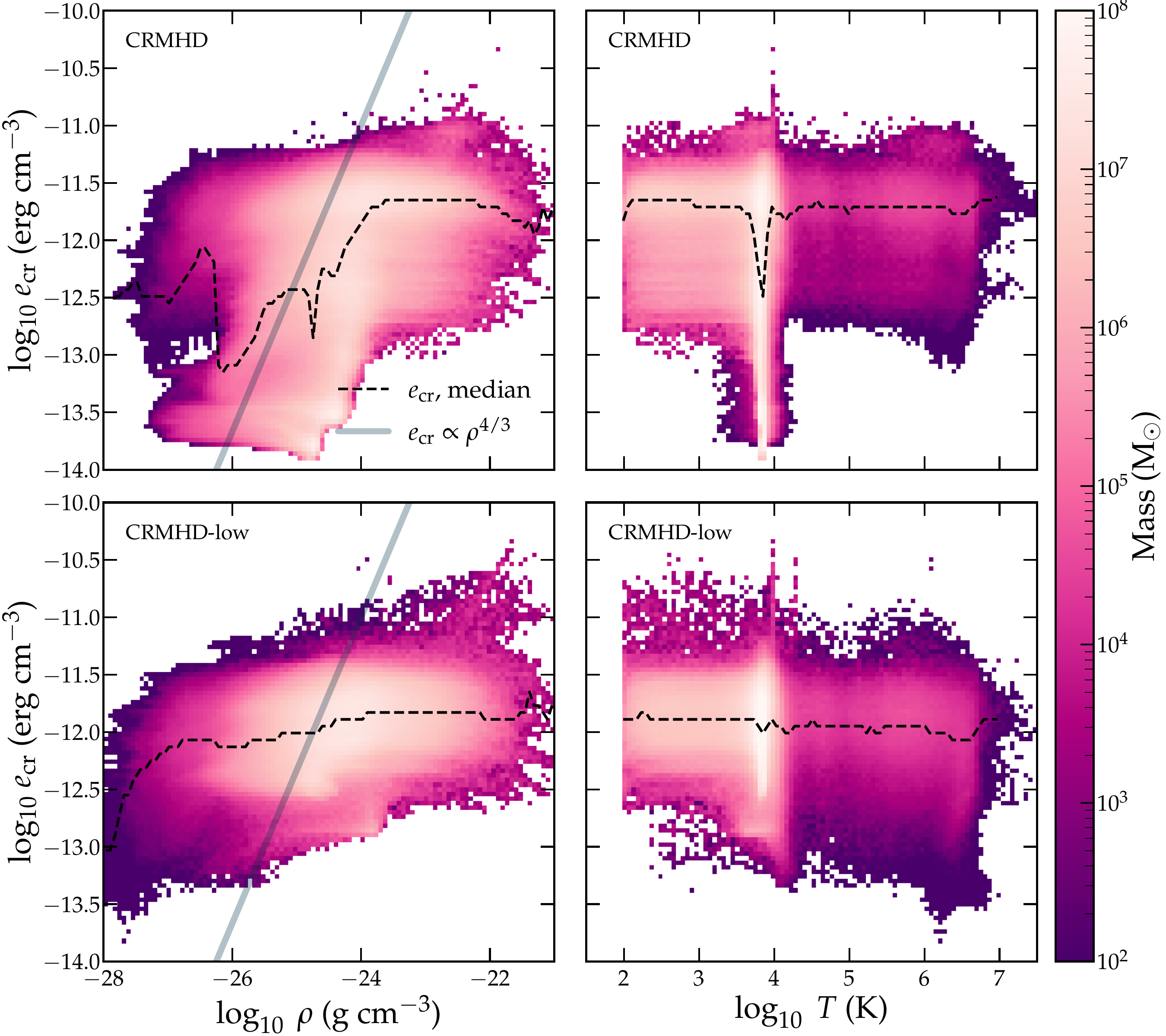}
    \caption{Distribution of the CR energy density as a function of density (left column) and temperature (right column) at $t=1.5$ Gyr, weighted by the gas mass. Overlaid is the median CR energy density in each gas density bin (dashed line). Additionally, in the two leftmost panels, we show the analytical scaling that assumes adiabatic CRs (solid gray line).}
    \label{fig:ecr_hist}
\end{figure*}

\begin{figure*}
    \includegraphics[width=\textwidth]{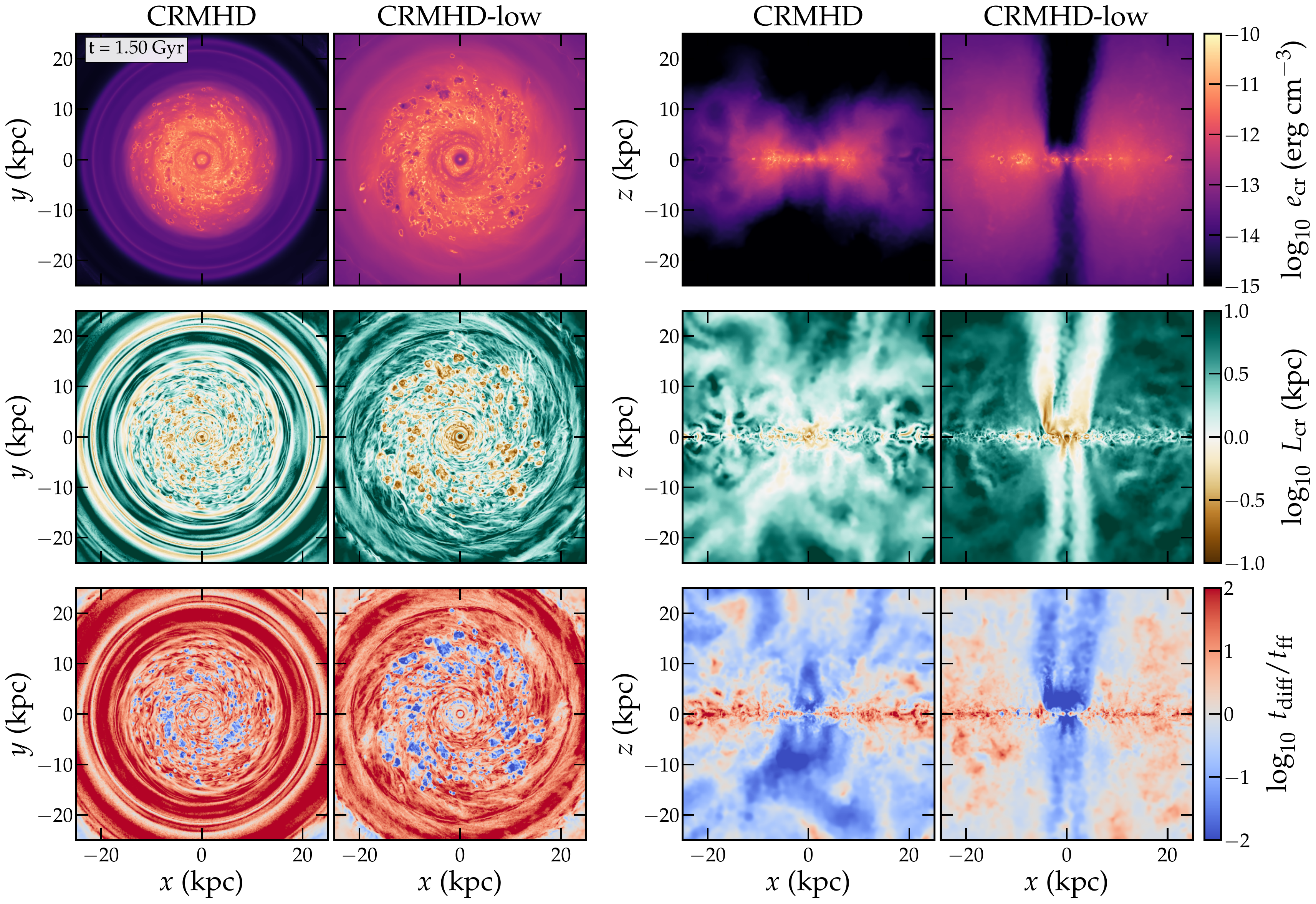}
    \caption{From top to bottom:\ Face-on and edge-on slices through the midplane of the CR energy density, $e_{\rm cr}$, the CR diffusion length, $L_{\rm cr}$, and the ratio of the CR diffusion timescale, $t_{\rm diff}$, to the free-fall timescale, $t_{\rm ff}$, for the CRMHD and CRMHD-low galaxies.}
    \label{fig:ecr}
\end{figure*}

\begin{figure*}
\centering
    \includegraphics[width=0.95\textwidth]{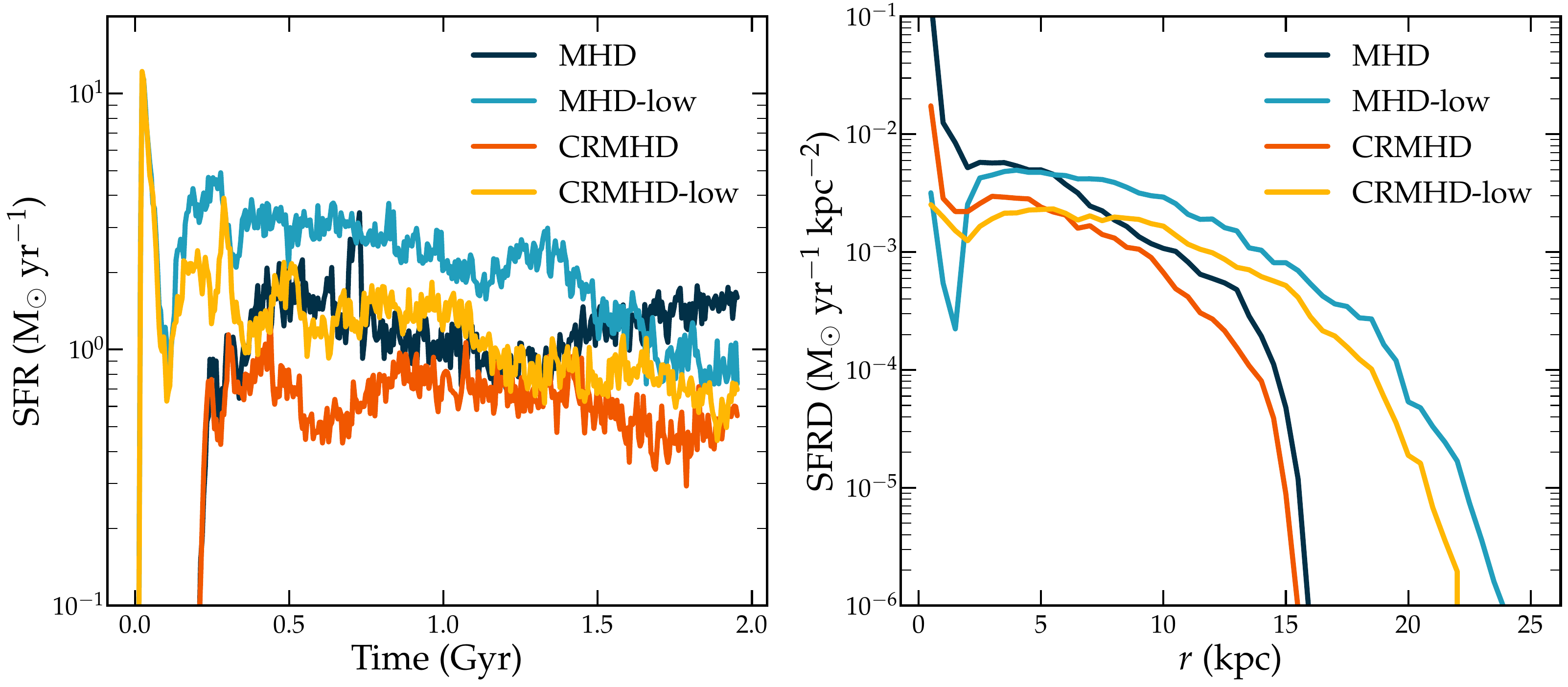}
    \caption{Left: SFR of our simulations as a function of time. Right: SFRD as a function of galactocentric radius, averaged over $t=0.5-2.0$ Gyr.}
    \label{fig:sfr}
\end{figure*}

To illustrate the gas phase distribution of the ISM in our simulations we show the density-temperature histograms in the bottom row of Figure \ref{fig:scaleheight_phaseplot}, taken at $t=1.5$ Gyr . The histograms are weighted by the gas mass and we have only considered gas in the disk. We note that the phase plots are visually remarkably similar between the simulations and show the different phases of the ISM we would expect to see. Between 70\% and 80\% of the mass is contained in the warm neutral medium in the temperature range $5050~\mathrm{K}<T<2\times10^{4}$~K \citep{Kim_2018}, with 20\%-30\% of the mass contained in the cold neutral medium at lower temperatures (consistent with observational data, see, e.g.,\ \citealt{Ferriere2001} or \citealt{Klessen2016}). The regions in the galaxies with very hot, low-density gas are recently formed SN bubbles. The CR-simulations are truncated at slightly larger gas densities than their MHD counterparts. For example, there is barely any gas with $\rho<10^{-28}\;\mathrm{g}\;\mathrm{cm}^{-3}$ in CRMHD-low, unlike in MHD-low. This is consistent with the somewhat smoother appearance of the CRMHD-galaxies we noted in the previous section. The CRs tend to wash out thermal pressure gradients as they diffuse out from their injection sites, resulting in fewer SNe exploding in low-density environments that create large, low-density bubbles.

\subsection{CR energy distribution}
In Figure \ref{fig:ecr_hist}, we illustrate how the CR energy density is distributed in our simulations CRMHD and CRMHD-low. In the left column we plot the CR energy density against gas density, while the right column shows the CR energy density distribution plotted against the gas temperature. All histograms are weighted by the gas mass and, once again, only gas in the disk is considered here. We also indicate the median CR energy density as dashed black lines. In the CRMHD simulation we have more low-energy CRs than in the CRMHD-low case, leading to a minimum in the median CR energy density at around $T=10^{4}$~K. This is because the star-forming disk is smaller in the CRMHD-simulation, meaning we are including some regions in the outskirts of the galaxy, where no star formation is taking place.

For adiabatic changes to the CR energy, due to either compression or expansion of the gas, we expect the CR energy density to scale with the gas density as $e_{\mathrm{cr}}\propto \rho^{4/3}$ (e.g.,\ \citealt{Girichidis2024}), where $\gamma=4/3$ is the adiabatic index of the CR fluid. We visualize this scaling in Figure \ref{fig:ecr_hist} using solid gray lines. In our CRMHD galaxy we see a positive correlation between the CR energy density and the gas density, which is somewhat similar to the analytical relation at lower densities; however, the distributions flattens at higher densities ($\rho\gtrsim 10^{-24}\;\rm g\; cm^{-3}$), completely deviating from the $\rho^{4/3}$ scaling. In both simulations there is a large spread around the median value. In the CRMHD-low galaxy there is a weak positive correlation, but the average CR energy density is overall remarkably flat with gas density and unlike the adiabatic scaling. This tells us that the CRs in our galaxies do not behave adiabatically and, instead, their collective thermodynamic properties are dominated by non-adiabatic interactions between the CRs and the thermal gas, such as hadronic losses. CR diffusion also has the effect of flattening the CR energy density distribution, particularly at higher densities.

We investigate this point further by calculating the CR diffusion length, defined as $L_{\rm cr}=e_{\rm cr}/|\boldsymbol{\nabla}e_{\rm cr}|$, as well as the CR diffusion timescale, defined as 
\begin{align}
    t_{\rm diff} = \frac{L_{\rm cr}}{v_{\rm diff}}.
\end{align}
The diffusion length can be thought of as the typical distance CRs have spread to. The CR diffusion speed $v_{\rm diff}$ is defined as in \cite{Girichidis2024} as
\begin{equation}
    v_{\rm diff} = -\kappa\frac{\boldsymbol{b}\boldsymbol\cdot \boldsymbol\nabla {P}_{\rm cr}}{{P}_{\rm cr}},
\end{equation}
where $\kappa$ is the CR diffusion coefficient that we have set to $4\times10^{28} \ \rm cm^2 \ s^{-1}$ in all our simulations with CRs. In the second and third rows of Figure \ref{fig:ecr}, we show face-on and edge-on slices through the midplane of $L_{\rm cr}$ and $t_{\rm diff}/t_{\rm ff}$ at $t=1.5$ Gyr, where $t_{\rm ff}=\sqrt{3\pi/32G\rho}$ is the free-fall time. The ratio $t_{\rm diff}/t_{\rm ff}$ therefore assesses how the timescale that CRs act on compares to the timescale on which dense structures collapse. We see from the figure that in most of the disk the diffusion length is around 1 kpc, which is much larger than the typical scale of molecular clouds, for example. The diffusion of CRs through the disk weakens the CR pressure gradients in the galaxy, resulting in large diffusion lengths after 1.5 Gyr of evolution. One exception is  SN-driven bubbles, where CRs have recently been injected and the gradients are still strong. The interpretation of the timescales requires a more careful explanation. A long diffusion timescale corresponds to small normalized CR gradients and, thus, small diffusive speeds. This suggests that CRs will behave adiabatically since local gas motions can exceed the diffusive speeds and create local CR enhancements or voids. However, the large diffusion coefficients quickly remove local CR overdensities as soon as they appear due to adiabatic contraction or expansion. As a result, we do not see an adiabatic behaviour of the CR fluid, which would scale with $\rho^{4/3}$, as seen in Figure~\ref{fig:ecr_hist} here as well as Figure 1 in \citet{Girichidis2024}. Effectively, the small CR gradients result in weaker CR forces compared to local thermal gradients, meaning that dense structures collapse faster than the timescales at which CRs act.

In the top row of Figure \ref{fig:ecr}, we also show face-on and edge-on cuts of the CR energy density at $t=1.5$ Gyr. In the cavities left by SNRs the CR energy density is low, which is an artifact of the CR energy injection scheme. As was clear in the column density maps of Figure \ref{fig:morphology}, the CRMHD-low simulation results in a larger star-forming disk. In the edge-on maps in the right panels, we can clearly see the effect of the lower initial magnetic field in the CRMHD-low galaxy. Due to the weaker magnetic field, the CRs have diffused far into the CGM\karin{. As we can see in Figure \ref{fig:morphology}, this occurs early on in the simulation ($t\lesssim0.5$ Gyr)}. We also note the existence of two cones of very low CR energy emerging from the center of the galaxy, where the CRs have been transported away. The outflows in our simulations are discussed in more detail in Section \ref{sec:outflows}.

\subsection{Star formation rate}

We compare the SFRs from our four simulations in Figure \ref{fig:sfr}. In the left panel, we show the time evolution of the total SFR for all of our simulations, while in the right panel, we show the SFR surface density (SFRD) as a function of galactocentric radius, averaged over $t=0.5-2.0$ Gyr. In the evolution of the SFR of the MHD and CRMHD runs, we can see the effect of the complete mass return in SNe (recall Section \ref{subsec:sim details} for further details), as the SFRs do not exhibit the decrease over time as would otherwise be expected if gas were to be continually depleted\karin{. An example of this can b seen in Fig. 8 in \cite{göller2024}}. The MHD-low and CRMHD-low galaxies start off with a strong initial burst of star formation and then a weakly decreasing SFR over the 2 Gyr, likely because over the evolution the magnetic field builds up to similar values as in the CRMHD and MHD cases. The stronger magnetic field in the CRMHD and MHD galaxies leads to a later onset of star formation, as the gravitational collapse of gas must fight the pressure provided by the magnetic fields. At all times the SFR of the simulations with CRs lies below that of the corresponding MHD simulations with the same initial magnetic field strength, showing that the additional pressure from CRs makes it more difficult to form stars. The global SFR in the MHD and MHD-low simulations averaged over $t=0.5-2.0$ Gyr is 1.2 and 2.1 $\rm M_\odot \, {\rm yr^{-1}}$, respectively. For CRMHD and CRMHD-low the average SFR in this time frame is 0.62 and 0.96 $\rm M_\odot \, {\rm yr^{-1}}$, respectively. \karin{These values are consistent with observations that suggest the SFR of the Milky Way lies between $1-2\rm \ M_\odot \ yr^{-1}$ (e.g.,\ \citealt{Licquia_2015,Elia_2022}), although the SFR in the CR-simulations fall on the lower end.}

A similar behavior can be observed in the SFR density, shown in the right panel of Figure \ref{fig:sfr}. The inclusion of CRs gives a lower SFRD at all radii, with the median SFRD decreasing by 64\% and 14\% in MHD and CRMHD and MHD-low and CRMHD-low, respectively. In the MHD and CRMHD galaxies, the SFRD drops off rapidly at a radius of about 15 kpc; whereas in the MHD-low and CRMHD-low galaxies we have star formation out to around 23 kpc, consistent with what we see in Figure \ref{fig:morphology}.

\begin{figure*}
    \includegraphics[width=\textwidth]{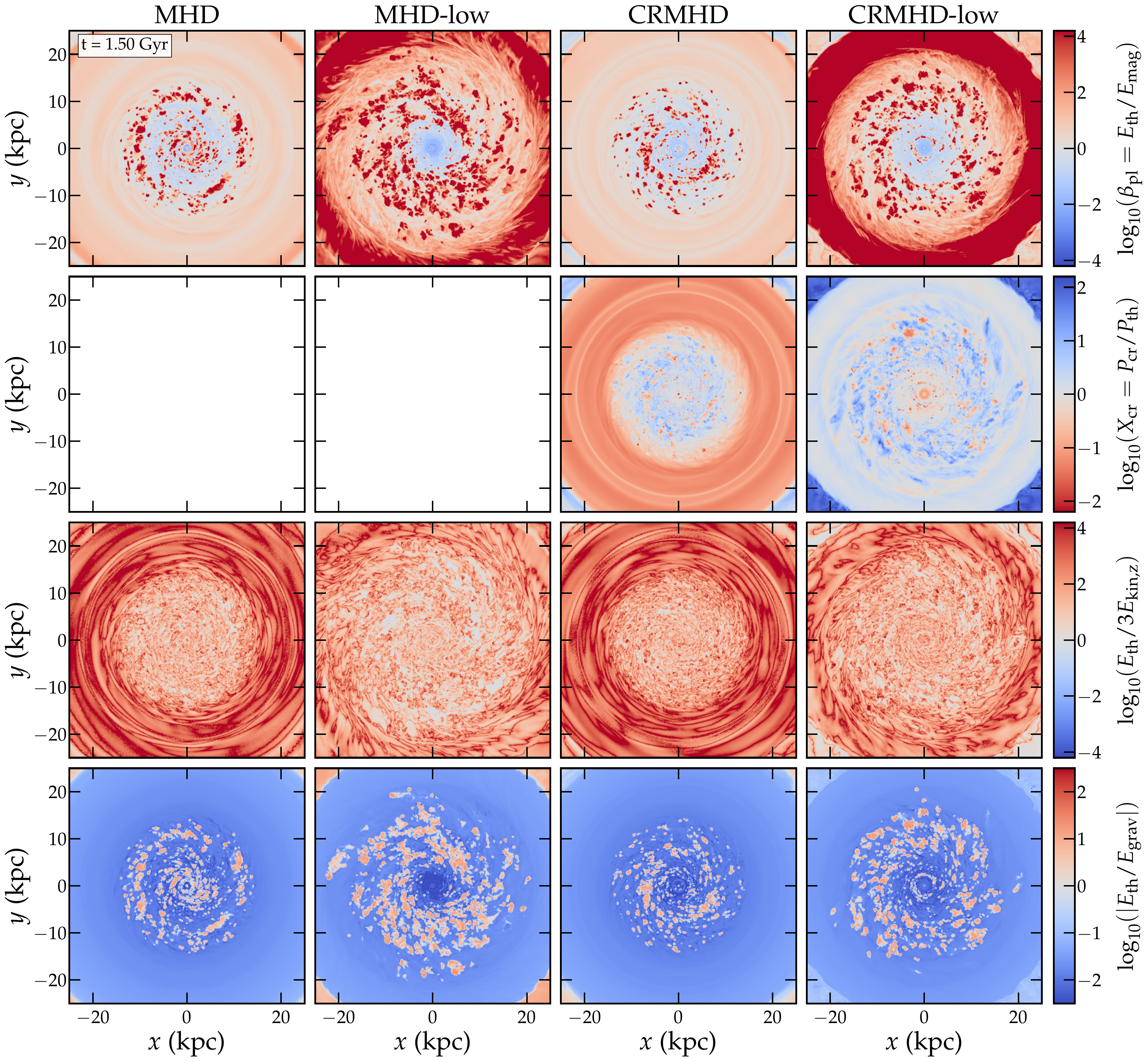}
    \caption{Slices through the galactic midplane of various energy ratios at $t=1.5$ Gyr. From top to bottom: Ratio of thermal to magnetic energy, ratio of CR to thermal pressure, ratio of thermal to the z-component of the kinetic energy, and the ratio of thermal to gravitational energy. In all maps, red signifies the dominance of thermal energy over the other component, while white regions denote regions with equipartition.}
    \label{fig:energies_faceon}
\end{figure*}

\begin{figure*}
\centering
    \includegraphics[width=0.95\textwidth]{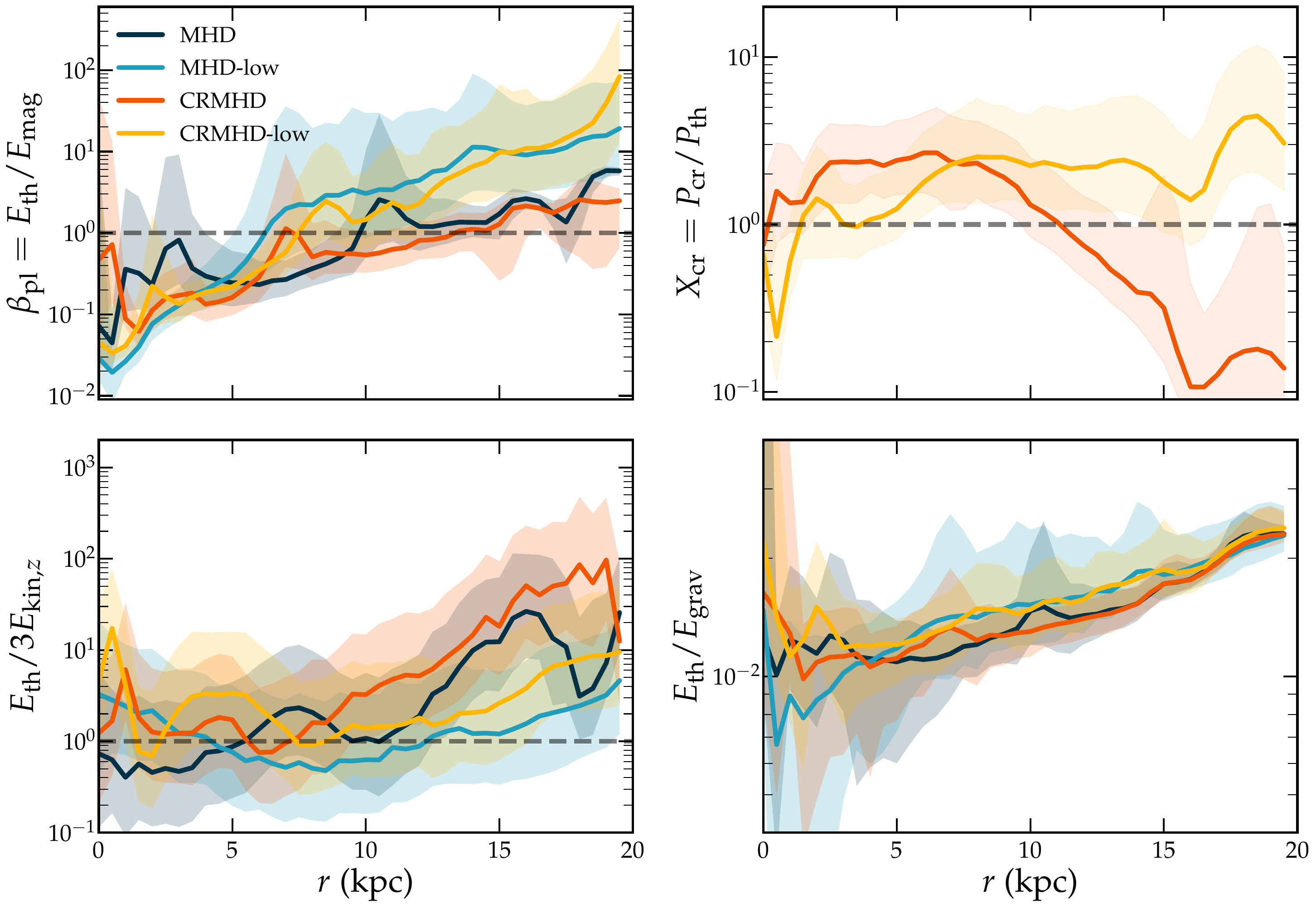}
    \caption{Radial profiles of the energy ratios from Figure \ref{fig:energies_faceon} in the galactic disk. The median values of the given energy ratio for each radial bin are plotted, considering cells within a height of 1 kpc and looking at our fiducial time $t=1.5$ Gyr. Shaded regions denote the 25th and 75th percentile. The dashed lines denote where the energy components are equal; i.e., where we have equipartition.}
    \label{fig:energies_rad}
\end{figure*}

\section{Energy distribution in the disk}\label{sec:energetics}
To examine the dominant energy components in our galaxies,  we used face-on cuts through the midplane of various energy ratios at $t=1.5$ Gyr in Figure \ref{fig:energies_faceon}. From left to right, we show our different simulations, and from top to bottom, we list the ratio of thermal to magnetic energy $\beta_{\rm pl}$, the ratio of CR to thermal pressure $X_{\rm cr}$, the ratio of thermal to turbulent kinetic energy, and  the ratio of thermal to gravitational potential energy. By convention, $X_{\rm cr}$ is defined as a pressure ratio so we keep it that way for easier comparison with other works, but we note that this is a factor of 2 smaller than the ratio $E_{\rm cr}/E_{\rm th}$ due to the different adiabatic indices. The kinetic energy in the $x-y$ plane is dominated by galactic dynamics such as rotation, so we used only the $z$-component to probe the turbulent kinetic energy. Under the assumption of isotropic turbulence, the energy in the different components should be the same, so that $E_{\rm kin,turb}=3E_{\rm kin,z}$. The gravitational potential energy is calculated by \textsc{Arepo} and includes contribution from self-gravity and the external potential. Redder areas in the figure denote regions where the thermal energy dominates over the other component and whiter areas denote regions where there is equipartition between the two energies. Additionally, in Figure \ref{fig:energies_rad}, we plot the radial evolution of these ratios for all our galaxies at $t=1.5$ Gyr. For these radial profiles, we considered cells with $h\leq1$ kpc, plotting only the median value in each radial bin. The dashed horizontal lines denote equipartition and the shaded regions denote the 25th and 75th percentile. 

The CR energy density is clearly an important energy component in the disk, as most of the disk appears blue or blueish in the face-on maps of $X_{\rm cr}$. Regions of presumably relatively recent SN explosions are thermally dominated, where the CR energy is 10\% of the thermal energy, as per our injection scheme. If we could resolve the expanding SNR shocks where the CRs are accelerated, we would likely find that CR pressure dominated over thermal pressure also inside the bubbles, as adiabatic expansion favors the CR pressure that has a softer equation of state \citep{Pfrommer2017}. Nonetheless, in most of the disk is CR energy dominates over the thermal, turbulent, and magnetic components. 

We note that in most of the disk the thermal and magnetic energy components are comparable, except in the hot, low-density bubbles created by SNe where the thermal energy naturally dominates. Surprisingly, it seems as if in the two galaxies with an initially lower magnetic field, MHD-low and CRMHD-low, by 1.50 Gyr of evolution the magnetic energy is stronger in the center compared to the thermal component than in the MHD and CRMHD runs. \karin{This could be due to more efficient magnetic field amplification in the centers of the CRMHD-low and MHD-low galaxies as a result of the initial phase with stronger star formation activity (see Figure~\ref{fig:sfr}).} Looking at the radial evolution, we see that the thermal component increases in dominance as we move further out in the disk, but we have equipartition within one order of magnitude. 

The thermal and turbulent kinetic energy are generally compatible with no distinctive regions where one or the other dominates. The gravitational potential energy clearly dominates over the thermal component, as we would expect for disk galaxies where the gravitational energy is largely balanced by kinetic energy from the rotation of the galaxy. In the recently exploded SN bubbles, however, the thermal energy weakly dominates. This indicates that SN bubbles are able to expand against gravity and eventually break out of the midplane, generating fountain flows. There is no major difference between the simulations for these particular energy ratios. The properties of outflows generated in our simulations is discussed in the next section.

\begin{figure*}[h!]
    \includegraphics[width=\textwidth]{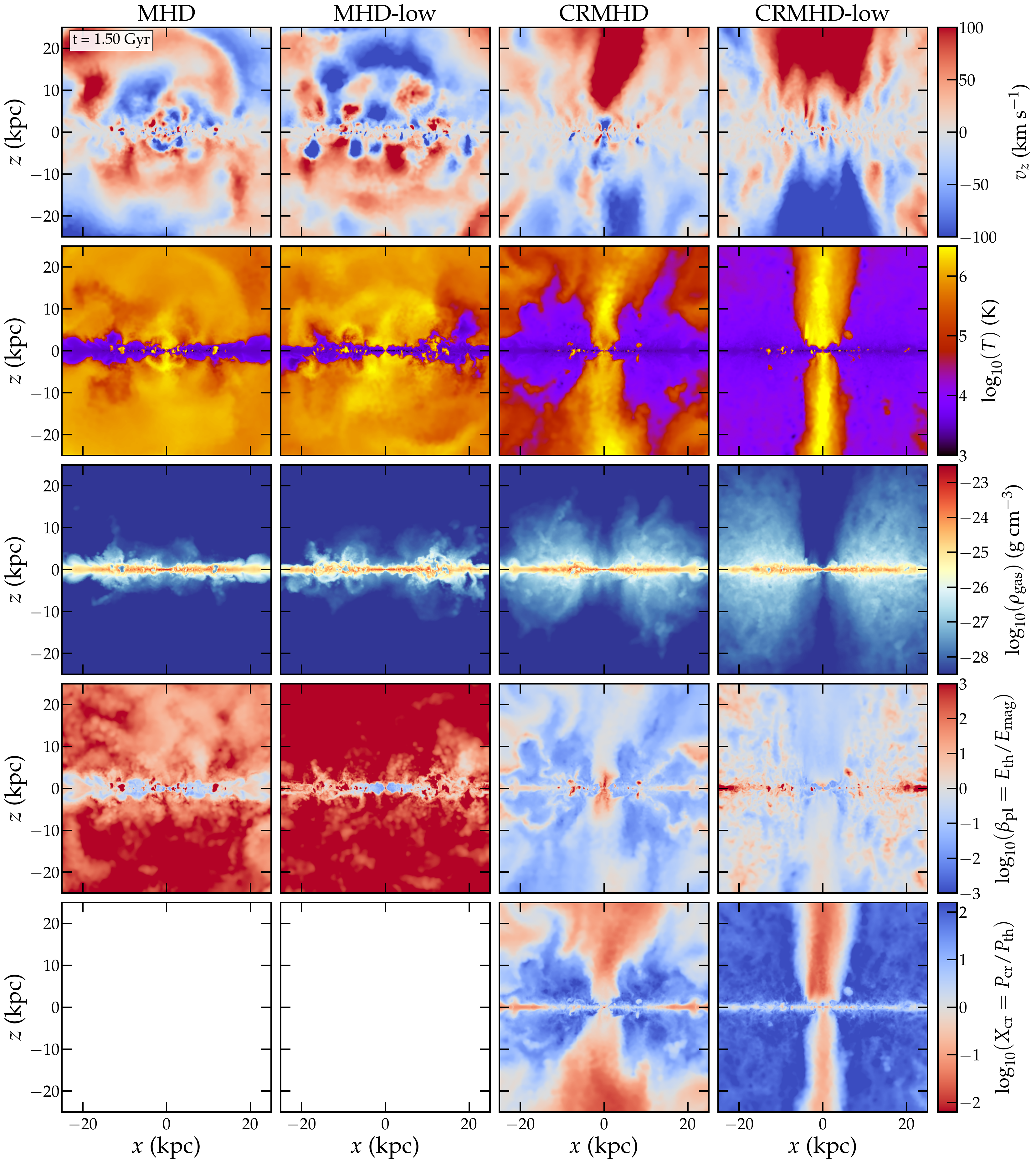}
    \caption{Comparison of the vertical structure and CGM properties of our galaxies at $t=1.5$ Gyr. From top to bottom, we show slices through the midplane of the vertical velocity,  gas temperature,  gas density,  magnetic-to-thermal energy ratio, and the CR-to-thermal pressure ratio. Without CRs, we would have a turbulent, hot CGM with a mix of inflow and outflow. CRs help launch hot, thermally dominated outflows from the center and mix colder gas into the CGM.}
    \label{fig:edge_maps}
\end{figure*}

\begin{figure*}
\centering
    \includegraphics[width=0.9\textwidth]{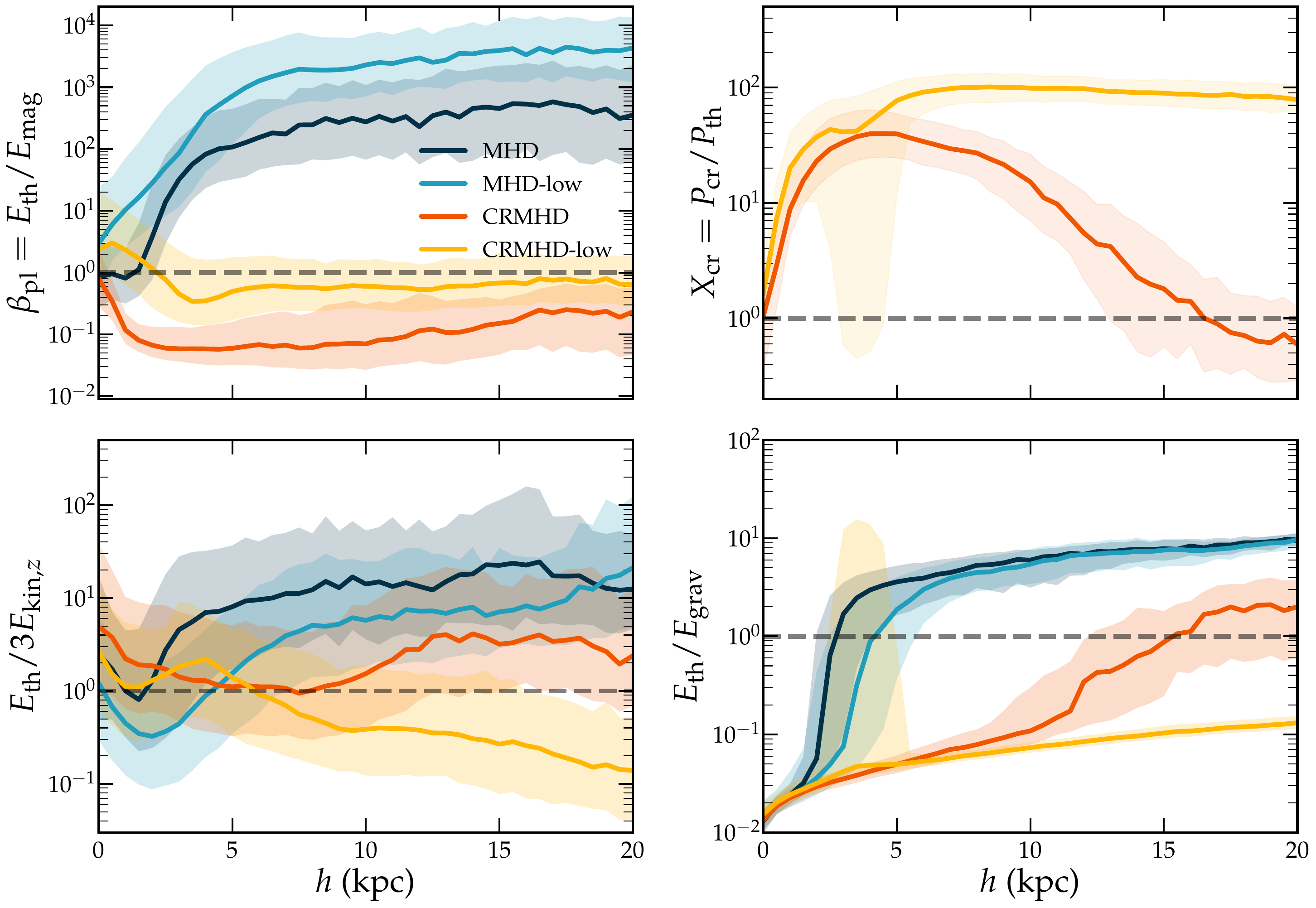}
    \caption{Same as Figure \ref{fig:energies_rad}, but showing the vertical profiles, averaged within a cylinder of $r=\rm 20$ kpc. Once again, we are plotting the median values of each height bin, with the shaded areas denoting the 25th and 75th percentile.}
    \label{fig:energies_vert}
\end{figure*}

\begin{figure*}
    \includegraphics[width=\textwidth]{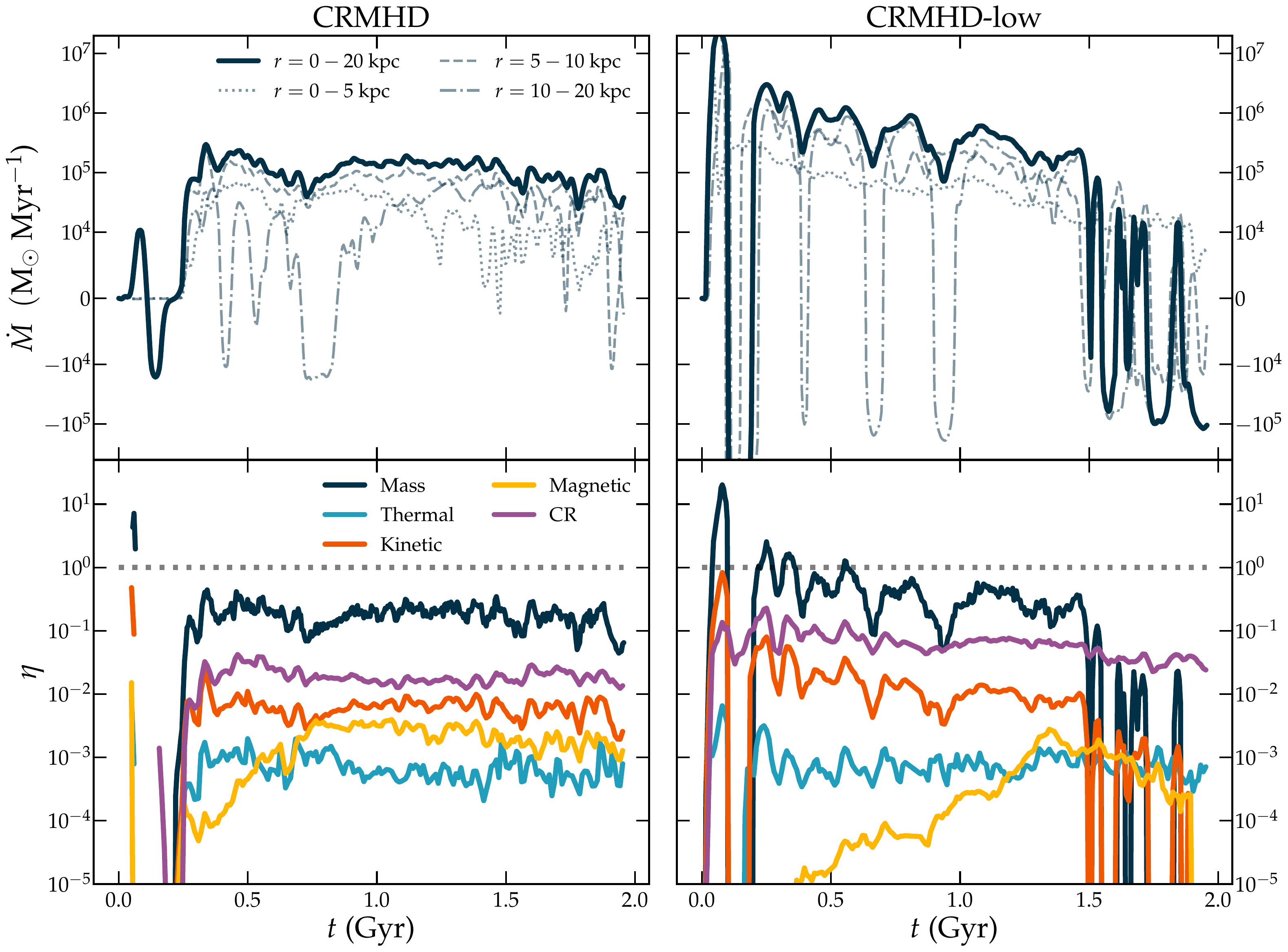}
    \caption{
    Evolution of the mass flow rate.
    In the two upper panels we show the evolution of the mass outflow rate (mass flux) in the vertical direction at a height of $h = 5$ kpc in different radial bins, denoted by different linestyles, for the simulations CRMHD and CRMHD-low. Negative values correspond to net inflow. In the lower panels we show the mass-loading factor and energy loading factors, defined in Eqs.\ \eqref{eq:edot_kin}-\eqref{eq:edot_mag}\karin{, calculated at the same height as the mass outflow rates and for radii $r<20$ kpc.}}
    \label{fig:outflows}
\end{figure*}

\begin{figure}
    \includegraphics[width=0.94\columnwidth]{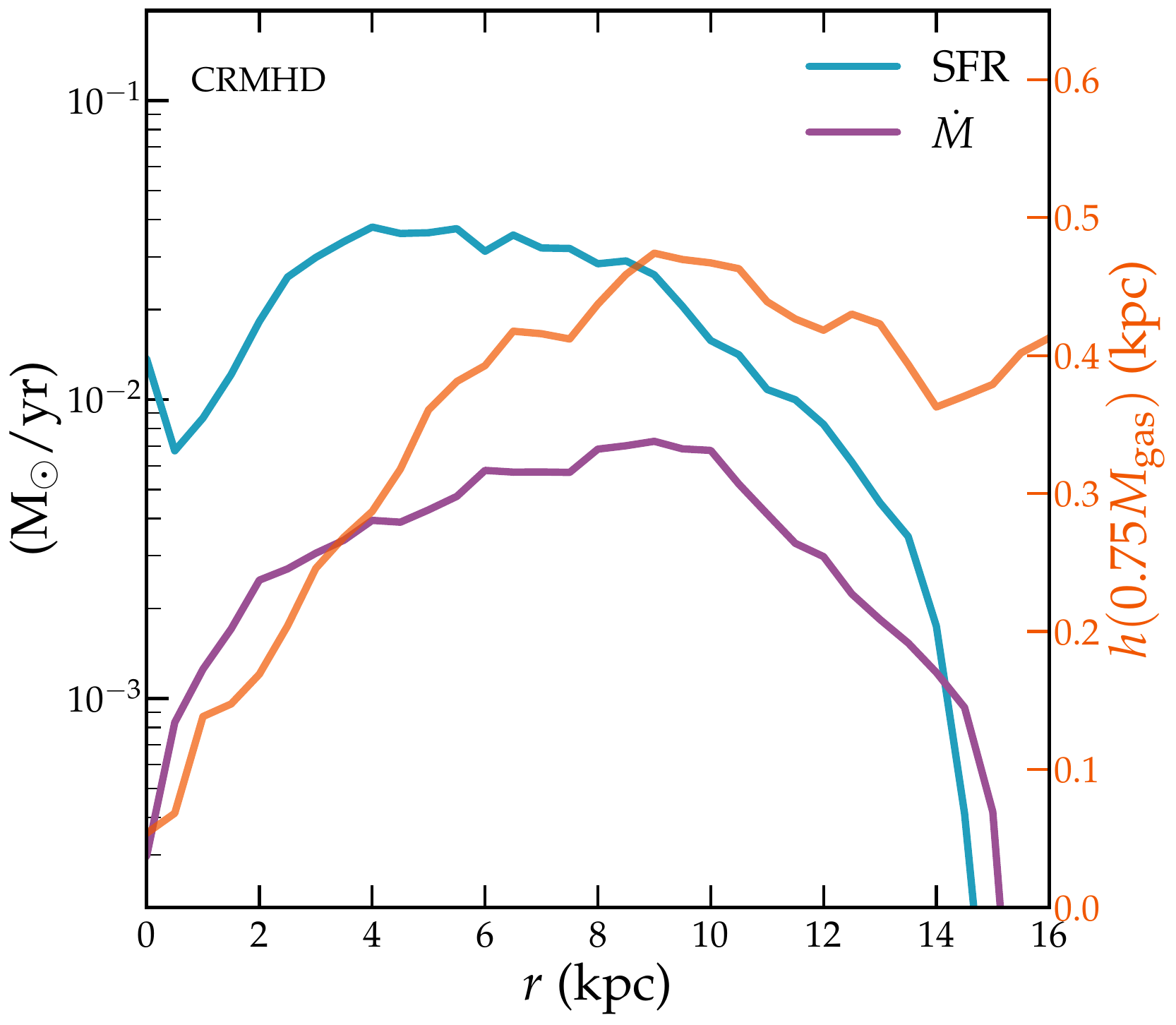}
    \caption{Radial evolution of mass outflow rate, $\dot{M}$, and SFR for the CRMHD simulation, averaged over $t = 0.5-2.0$ Gyr. 
    The pink curve shows the gas scale height, $h$, averaged over the same time interval. Its values at the largest $r$ imply that CRs lift gas at all radii, not just in the center.
    }
    \label{fig:outflows_rad}
\end{figure}

\begin{figure}
    \centering
    \includegraphics[width=\columnwidth]{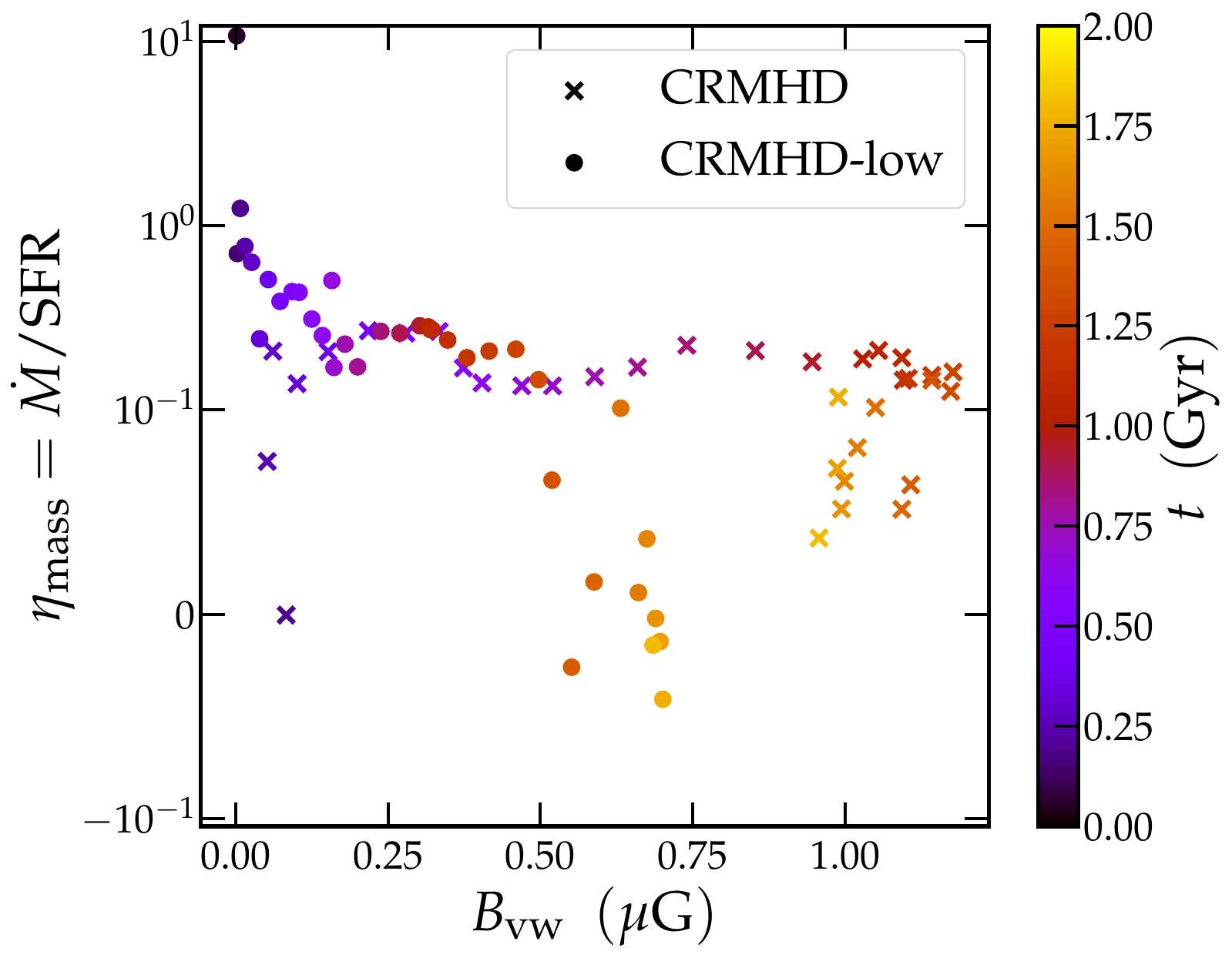}
    \caption{Mass-loading factor as a function magnetic field strength at a height of $5\,\mathrm{kpc}$ for both CR simulations. Colour coded is the simulation time. At early times the simulations differ. At late times the mean field strengths are comparable and so are the mass-loading factors.}
    \label{fig:mass-loading-Bfield-CGM}
\end{figure}

\begin{figure*}
\centering
    \includegraphics[width=0.75\textwidth]{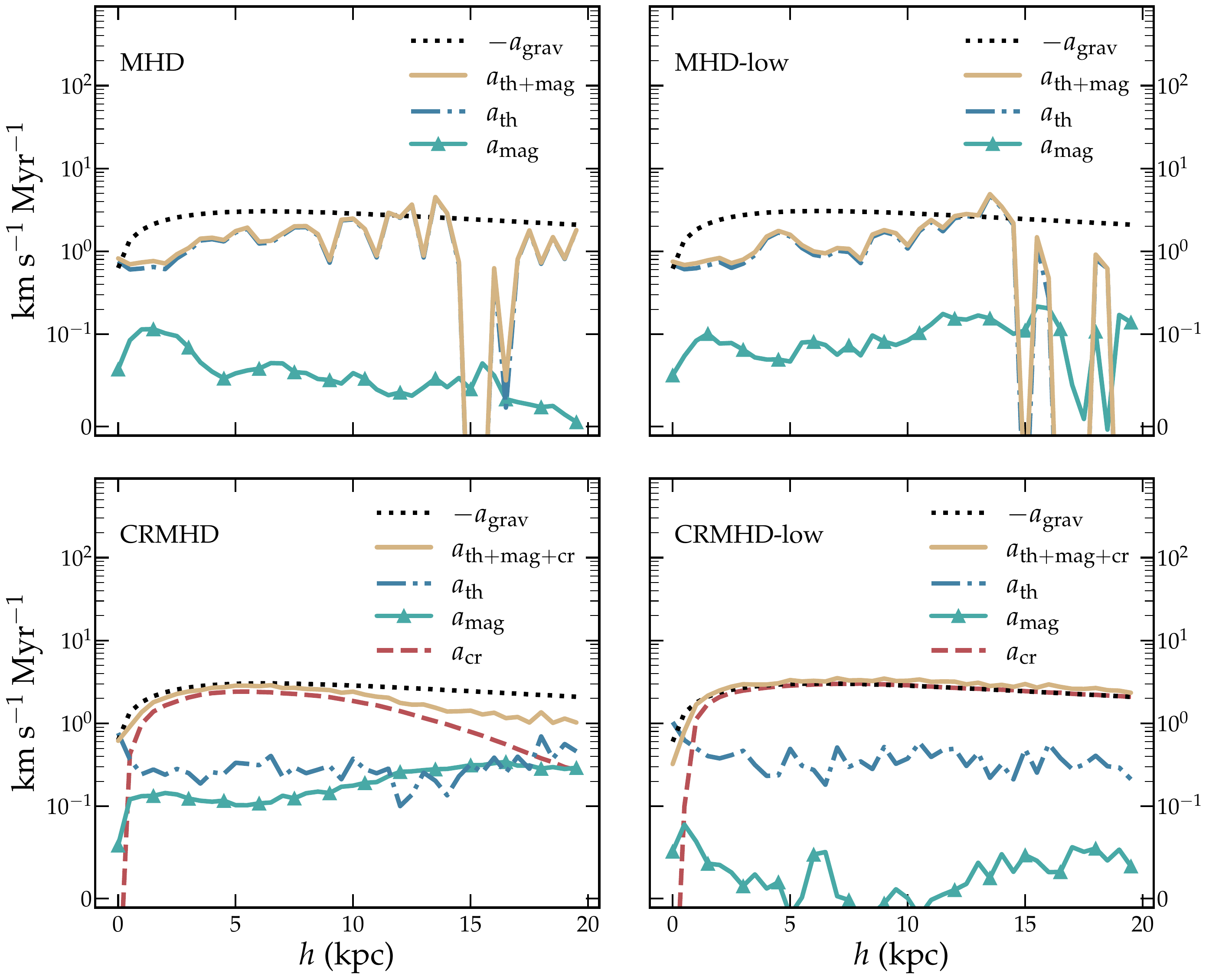}
    \caption{Time-averaged ($t = 1.25-2.00$\,Gyr) vertical accelerations within a cylinder of $r=20$\,kpc for the four simulation setups. Accelerations are first volume averaged and then averaged in time. The pure MHD-cases do not exhibit outflows, so the gravitational force dominates. In the CRMHD simulation, the CR acceleration dominates the outward forces and is comparable to the inward gravitational acceleration up to a height of around 10\,kpc. The CRMHD-low galaxy is dominated by CR acceleration \karin{at all heights}.}
    \label{fig:acc}
\end{figure*}

\begin{figure*}
\centering
    \includegraphics[width=0.8\textwidth]{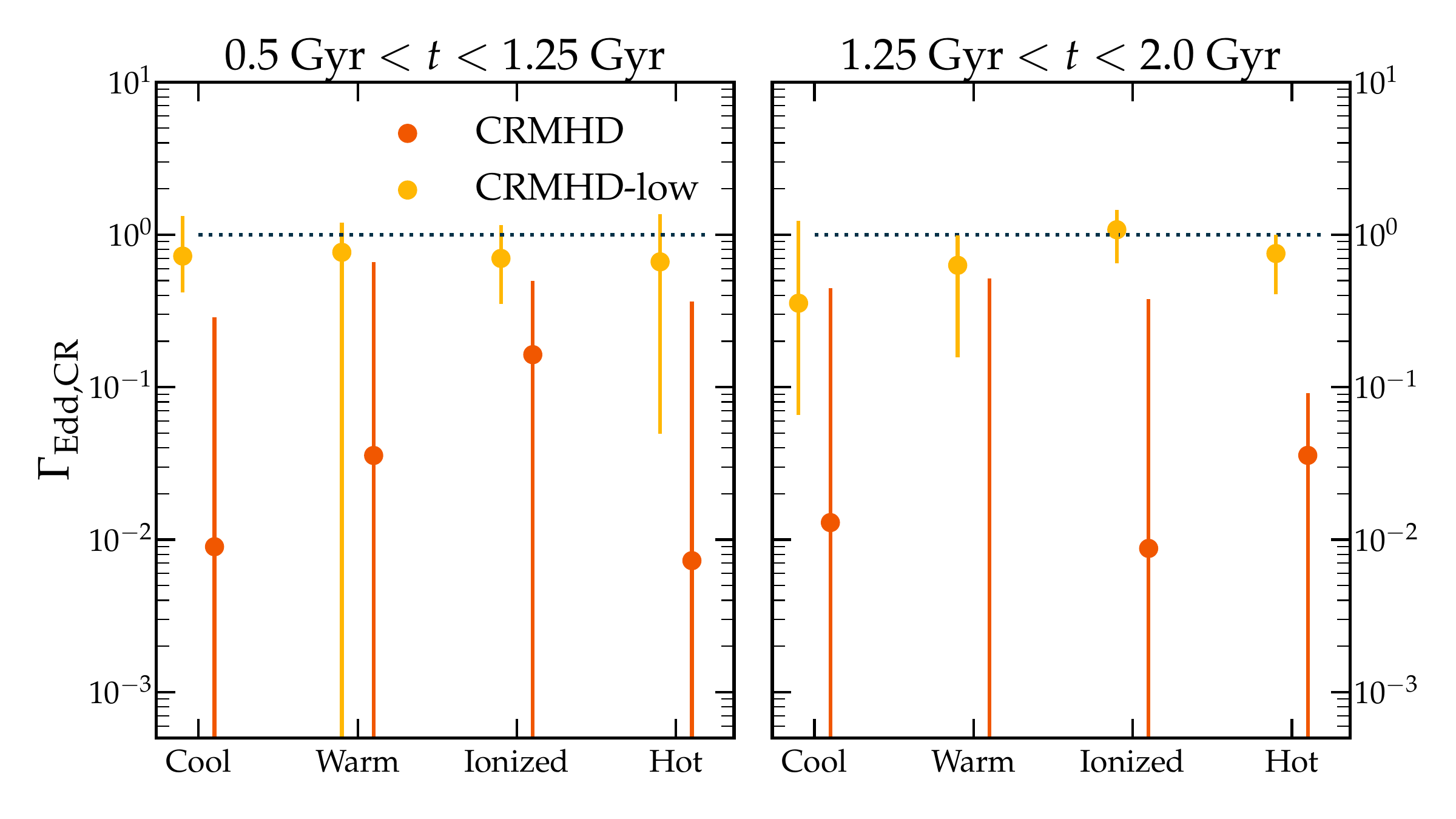}
    \caption{Volume-weighted CR Eddington factors averaged in a region $r<20$ kpc and $4< h <6$ kpc, separated by gas phase (cool phase: $T<5050$ K, warm phase: $5050\;\mathrm{K}<T<2\times10^4\;\mathrm{K}$, ionized phase: $2\times10^4\;\mathrm{K}<T<5\times10^5\;\mathrm{K}$, hot phase: $5\times10^5\;\mathrm{K}<T<10^{10}\;\mathrm{K}$). We plot the median value and the 25th and 75th percentile averaged over an early time interval ($t=0.5-1.25$ Gyr, left panel), and a later time interval ($t=1.25-2.0$ Gyr, right panel).}
    \label{fig:edd}
\end{figure*}

\section{Vertical structure and outflows}\label{sec:outflows}
In this section, we investigate the vertical structure of our galaxies in detail. In particular, we explore the outflow region.

\subsection{Vertical gas structure}
In Figure \ref{fig:edge_maps}, we show edge-on cuts through the midplane of various physical quantities at $t=1.5 \rm~Gyr$. Each column represents a different simulation. From top to bottom, we show cuts through the center of the vertical component of the velocity, $v_z$,  gas temperature, $T$,  gas density, $\rho_{\rm gas}$,  ratio of thermal to magnetic energy, $\beta_{\rm pl}$, and the ratio of CR to thermal pressure, $X_{\rm cr}$. In the MHD and MHD-low galaxies there is a mix of inflowing and outflowing gas, but no visible coherent outflows. In  MHD-low especially, we clearly see kpc-sized bubbles of fast outflowing material from the entire disk, which result from expanding SN-driven bubbles that have broken out of the midplane. These outflows soon fall back onto the galaxy as small fountain flows. The lack of proper outflows results in a hot, thermally dominated CGM, with a thin disk. In Figure \ref{fig:energies_vert}, we portray the same energy ratios as in Figure \ref{fig:energies_rad}, but now averaged in vertical bins ($r\leq20$ kpc) to quantify how energy is distributed in the CGM. In the MHD-galaxies the thermal energy  \karin{dominates over all other energy components in the CGM, in particular, the magnetic energy}.

When we include CRs we get cones of hot ($T>10^6$ K), low-density, fast-moving gas launched from the galactic center, with velocities exceeding 100 km/s. This is consistent with results from other hydrodynamical simulations that also find bipolar outflows driven by CR pressure \citep[e.g.,][]{pakmor2016,Girichidis2024}. Although powered by CRs, the cones are dominated by thermal pressure rather than CR pressure. They are also rich in magnetic energy despite being full of hot gas and very few CRs. The magnetic field tends to align with the outflow cones, resulting in the CRs diffusing very rapidly in that direction. Apart from the outflow cones, the CR-galaxies also develop fountain flows and even directly beneath the cones, there is inflowing gas. Overall, the CGM is denser than in the MHD case, with a smooth density distribution, in particular for the CRMHD-low galaxy. However, as we see in Figure \ref{fig:scaleheight_phaseplot}, the scale heights of the four galaxies are very similar, implying that the gas pushed out into the CGM of the CR-simulations is of low mass, compared to the mass of the disk.

The inclusion of CRs leads to a strongly magnetized CGM, as outflows driven by the CRs break out magnetic field lines out of the disk, lifting magnetic flux to larger vertical heights \citep[e.g.,][]{HanaszStrongGirichidis2021}. Even when the galaxy was initialized with a very weak magnetic field, as in CRMHD-low, the magnetic energy density is comparable to the thermal energy density in practically the entire CGM. The difference in $\beta_{\rm pl}$ in the gas above and below the galactic plane between the CR- and MHD-runs is clearly visible in Figure \ref{fig:energies_vert}, where we have $\beta_{\rm pl}=10^{-1}$ and $\beta_{\rm pl}=1$ in CRMHD and CRMHD-low, respectively. This is between three and four orders of magnitude lower than in the galaxies without CRs.

The CRs help lift the gas in the entire disk, and subsequent fountain flows help with mixing of the gas, resulting in a much colder CGM than in the pure MHD-case. In this colder, denser gas we are completely dominated by CR pressure, which supports the gas against condensing and falling back onto the disk. This behavior is consistent with results from recent studies which also find that CR feedback leads to cooler CGMs dominated by CR pressure \citep{Buck_2020,Ji_2020,DeFelippis_2024,Girichidis2024}. In the vertical profile of the CR to thermal pressure in Figure \ref{fig:energies_vert}, we see that in the CRMHD galaxy the ratio $X_{\rm cr}$ peaks at around 5 kpc above the midplane, with CR pressure almost 100 times that of the thermal pressure. The ratio decreases as we move further away from the center, approaching equipartition at around $h=20$ kpc. In CRMHD-low on the other hand the CR to thermal pressure ratio is approximately constant up to this height.

\subsection{Outflow rate and loading factors}
In Figure \ref{fig:outflows}, we quantify the outflows in our galaxies with CRs. In the two MHD-galaxies that do not contain CRs, we do not see any coherent outflows, as the thermal injection from SNe is not enough to continually lift gas, so they are not pictured here. In the two upper panels, we show the mass outflow rate in four different radial bins as a function of time. The outflow rate is analyzed in slices parallel to the midplane at $|z|=5$ kpc. We calculate the mass outflow rate as 
\begin{align}
    \dot{M}=\sum_{\mathrm{cells} \ i}\rho_i v_{z,i} \mathrm{d}A,
\end{align}
where $\rho_i$ is the gas density of the cell in the region, $v_{z,i}$ is the vertical velocity component, and $\mathrm{d}A$ is a small area element. As we see in Figure \ref{fig:scaleheight_phaseplot} the galactic disks in all our galaxies are mostly contained within 1 kpc of the midplane, meaning that gas moving at a height of 5 kpc can be classified as outflows. 

In the CRMHD simulation, which has a stronger initial magnetic field, we initially do not see a net outflow. At around $t=0.25$ Gyr, we start to see coherent outflows of $10^5$ $\rm M_\odot / \rm Myr$ in the center ($r \leq 5$ kpc) and in the intermediate radial bin ($r=5-10$ kpc). Around $t=1.0$ Gyr the outflow in the center decreases, but we start to see outflows in the outermost radial bin ($r=10-20$ kpc). In CRMHD-low, we initially get a strong burst of outflows, which matches the initial burst in star formation seen in Figure \ref{fig:sfr}. We also see consistent outflows in all radial bins, with some moments of infall in the $r=10-20$ kpc bin. After $t\approx1.5$ Gyr, the evolution of the mass outflow rate changes, as the outflows at all radii disappear, except in the very center. At this point, the CRMHD-low galaxy has developed kpc-sized perturbations above and below the disk, resulting in inflow at this height. However, at larger heights above the midplane the outflows are still coherent. To illustrate this, we show outflow rates at $|z|=10$ kpc in Appendix \ref{app:outflows_10kpc} (Figure \ref{fig:outflows10kpc}).

During the evolution of CRMHD and CRMHD-low, we see coherently sustained outflows both in the center and further out in the disk, evident from the different radial bins portrayed in Figure \ref{fig:outflows}. To further visualize this, we show a radial profile of the mass outflow rate in the CRMHD run in Figure \ref{fig:outflows_rad}, together with the SFR and the average scale height. The quantities are averaged over the time interval $t=0.5-2.0$, using every tenth snapshot. The outflow rate peaks at around $r=9$ kpc, close to the solar circle, but is distributed smoothly over the entire disk.

We also investigated the outflows in CRMHD and CRMHD-low in terms of mass- and energy-loading, which is depicted as functions of time in the lower panels of Figure \ref{fig:outflows}. The mass-loading factor, $\eta_M$, is a way to quantify the efficiency of stellar feedback in launching outflows, and is defined as $\eta_M=\dot{M}/\rm SFR$, where we average the SFR over 10 Myr periods. 

We find average mass-loading factors of  $\sim$0.18 and $\sim$0.25 in the CRMHD and CRMHD-low simulations, respectively. As we did not expect to see any strong outflows in Milky Way-like galaxies, this is not surprising (see Section \ref{sec:discussion} for further comparison with numerical and observational work on outflows in Milky Way-like galaxies). The differences between CRMHD and CRMHD-low are subtle and depend on the evolutionary stage of the simulation. In Figure~\ref{fig:mass-loading-Bfield-CGM}, we highlight the time evolution of the mass-loading factor for the two CR models. The simulation with weak initial magnetic field (CRMHD-low, circles) does not provide any resistance in the CGM, which leads to fast and powerful outflows at early times with mass-loading factors on the order of 10. After a simulation time of $\sim1\,\mathrm{Gyr,}$ the field in the CGM has reached comparable strengths for both models and $\eta_M$ converges to similar values of order $0.2$. At late times, the outflows stall and locally even start falling back onto the disk. At this point both simulations have comparable field strengths of 0.7 (CRMHD-low) and 1 $\mu\mathrm{G}$ (CRMHD) at the measurement height of $5\,\mathrm{kpc}$.

The energy loading factors similarly assess how efficiently energy is transported in the outflow and are defined as
\begin{equation}
    \eta_E=\frac{\dot{E}}{10^{51} \ \rm erg \cdot SNr}, 
\end{equation}
where SNr is the SN rate in $\rm s^{-1}$ and $\dot{E}$ is the outward energy flux of the different energy components: thermal, kinetic, magnetic, and CR. The energy fluxes are defined as
\begin{align}
    \dot{E}_{\mathrm{kinetic}}&=\sum_{\mathrm{cells} \ i}\frac{1}{2}\rho_i v_i^2v_{\mathrm{out},i}\mathrm{d}A\label{eq:edot_kin},\\ 
    \dot{E}_{\mathrm{thermal}}&=\sum_{\mathrm{cells} \ i}(e_{\mathrm{th},i}+P_{\mathrm{th},i})v_{\mathrm{out},i}\mathrm{d}A,
\end{align}
\begin{align}
    \dot{E}_{\mathrm{CR}}&=\sum_{\mathrm{cells} \ i}(e_{\mathrm{cr},i}+P_{\mathrm{cr},i})v_{\mathrm{out},i}\mathrm{d}A\\
    &-\kappa{b}_{z,i}(\boldsymbol{b}_i\boldsymbol\cdot\boldsymbol\nabla e_{\mathrm{cr},i})\mathrm{sign}(z_i)\mathrm{d}A,\nonumber\\
    \dot{E}_{\mathrm{magnetic}}&=\sum_{\mathrm{cells} \ i}\frac{\boldsymbol{B}_i^2}{8\pi} v_{\mathrm{out},i}\mathrm{d}A, \label{eq:edot_mag}
\end{align}
where $v_{\mathrm{out},i}= v_{z,i}\,\mathrm{sign}(z)$. The CR energy flux includes an additional term that accounts for the transport of CRs relative to the gas, along the magnetic field and down the CR pressure gradient. We see in Figure \ref{fig:outflows} that the CRs dominate the energy loading budget of the outflows, with median values of $\sim$0.018 in the CRMHD case and $\sim$0.058 in the CRMHD-low case. Since we injected CR energy in SNe with an efficiency of 10\%, this means that 18\% and 58\% of the injected CR energy is transported out by the outflow in CRMHD and CRMHD-low, respectively. This indicates that in the strong-$B$ case, most CRs remain within the galaxy, losing energy through interactions with the gas. In the low-$B$ model most of the CRs escape galaxy, and the resulting efficiency is more comparable to the 80\% found in the simulations of \cite{Thomas2024}.

Aside from the CR energy, the outflows are dominated by kinetic energy rather than thermal energy. This is reasonable considering that the CR-driven outflows consist of colder ($T\sim10^4$~K, see \karin{Figure} \ref{fig:edge_maps}) gas. A non-negligible contribution to the energy loading budget in the CRMHD galaxy comes from the magnetic energy, which builds up over the 2 Gyr in the CRMHD-low case until it reaches similar values to the CRMHD simulation at around $t\sim1.4$ Gyr. The CGM becomes significantly more magnetized due to the presence of CRs, as we see in \karin{Figure} \ref{fig:edge_maps}, so we would expect the outflows to carry magnetic energy with them.

\subsection{Vertical acceleration}
In this section, we investigate how CRs contribute to the acceleration of gas by characterizing our outflows in terms of different force components (thermal, gravitational, magnetic, and CRs). We additionally analyze how CR acceleration affects different phases of the ISM by computing CR Eddington factors.

Figure \ref{fig:acc} depicts vertical acceleration profiles for our four simulations, using the same definition for the thermal, CR, and magnetic accelerations as in \cite{Girichidis2024}. The accelerations are computed for every tenth snapshot as the volume weighted average in a disk of radius $r=20$ kpc and thickness $\Delta z=0.5$ kpc. Since our goal is to investigate the long-term evolution, our accelerations are averaged over the time interval $t=1.25-2.00$ Gyr. Each panel corresponds to a different simulation. The total outward-pointing acceleration is shown as the solid gold line. We also show the individual force components; specifically: the thermal (dot-dashed line), magnetic (triangular markers), and the CR acceleration (dashed line). We also plot the negative gravitational acceleration in black. 

In the two MHD-galaxies (first two rows), the outward acceleration is completely set by the thermal acceleration. At nearly all heights, the gravitational attraction completely dominates over the outward forces. In all galaxies, the magnetic acceleration is negligible and is not affected by whether we initialize the simulation with a stronger or weaker magnetic field. Once we also include CRs (bottom two rows), this becomes the dominant force. In our CR-galaxy with a strong initial magnetic field (CRMHD), the CR acceleration closely matches the gravitational inward acceleration, but does not manage to overcome it. At heights greater than 10 kpc, the CR acceleration decreases rapidly. In CRMHD-low the vertical CR acceleration is stronger and is able to compensate for the gravitational attraction at all heights, except in the innermost $\sim$1 kpc.

Figure \ref{fig:edd} shows the volume-weighted CR Eddington factors for the CRMHD and CRMHD-low runs in different ISM phases, where the CR Eddington factor is defined as $\Gamma_{\rm Edd,CR}\equiv-a_{\rm cr}/a_{\rm grav}$. We computed volume-weighted averages in the region $r<20$ kpc and $4<h < 6$ kpc and then averaged over an early ($t=0.5-1.25$ Gyr) and a late ($t=1.25-2.00$ Gyr) time interval. The error bars in the figure correspond to the 25th and 75th percentile of this time-averaging. We show the CR Eddington factors in the cool ($T<5050$ K), warm ($5050\;\mathrm{K}<T<2\times10^4\;\mathrm{K}$), ionized ($2\times10^4\;\mathrm{K}<T<5\times10^5\;\mathrm{K}$), and hot ($5\times10^5\;\mathrm{K}<T<10^{10}\;\mathrm{K}$) phase. The definitions of the gas phases come from \cite{Kim_2018}, where we  combined the cool and cold phase, since we do not have many cold cells at these heights due to our resolution. 

In the CRMHD-low case, all phases are more efficiently accelerated than in the CRMHD case and we do not see a significant difference in the CR Eddington factor between the different phases. In the CRMHD case, we do see some variation. In the earlier time interval the warm and ionized phase are more efficiently accelerated and the CRs are less suited for accelerating the hot gas, similarly to the results of \cite{Sike2024}. At later times, the acceleration of the warm phase from CRs is actually negative at this height, since it is in this phase that we eventually see fountain flows.

\section{Discussion}\label{sec:discussion}

\subsection{Comparison to other simulation works and observations}

Several previous numerical works have explored the dynamical impact of CRs in Milky Way-like galaxies with setups resembling ours, albeit with notable differences. Furthermore, observations of outflows in the Milky Way and similar galaxies provide further context. In this section, we compare our results to such previous investigations.

\cite{Thomas2024}, used a similar setup of an isolated Milky Way-like galaxy, but with a two-moment CR transport. They found that CRs successfully power outflows, but the outflows in their MHD setups still reach mass-loading factors of $0.1-0.2$. Also \cite{montero2023}, with their cosmological simulations of Milky Way galaxies, reported stronger outflows in the MHD galaxies than those in ours; with the inclusion of CRs, they achieve mass loading exceeding unity. This is unlike our simulations, were we found weak outflows in the simulations with CRs and negligible mass-loading in the no-CR runs.

Our galaxies do not exhibit strong outflows. Even with the inclusion of CRs, we get global outflow rates of $0.1-1 \ \rm M_\odot\;\mathrm{yr}^{-1}$ and mass-loading factors of around $0.2$. Observational studies of star-forming galaxies have seen that for a fixed SFR, the outflow rate decreases with increasing stellar mass, so that higher mass galaxies are less effective at launching outflows due to their deeper gravitational potentials \citep{cicone_2016,JacobEtAl2018,Girichidis2024}. Indeed, estimations of the outflow rate in the Milky Way by \cite{Fox_2019} using UV absorption in high-velocity clouds (HVCs) yield an outflow rate of $\rm \dot{M}_{out}=0.16\pm 0.1 \ \rm M_\odot\;\mathrm{yr}^{-1}$ and a mass-loading factor of $\eta=0.10\pm0.06$. HVCs are clouds with velocities (relative to the local standard of rest) of $v>90\;\mathrm{km}\;\mathrm{s}^{-1}$, meaning that they are traveling too fast to be corotating with the galactic disk; therefore, they must represent inflows or outflows. Identical analyses of inflowing HVCs seem to actually imply that the Milky Way is in an inflow-dominated phase. Observations of other star-forming galaxies of comparable SFRs and stellar masses to the Milky Way have yielded similarly low outflow rates and mass-loading factors \citep[e.g.,][]{Roberts_Borsani_2020}. The outflow rates we find in our \karin{CR-}simulations are consistent with these observed values, although a caveat is needed since the way outflow rates are calculated in observations versus simulations differs fundamentally. Observational values provide instantaneous measures and may not represent typical outflow rates. Additionally, the nature of CR-driven outflows makes them inherently difficult to observe, as we expect a steady, slow lifting of gas at all radii. Outflowing low-velocity gas is unobservable in UV absorption due to blending with ISM foreground absorption. Furthermore, the degree of clumping in the gas has a significant impact on the derived outflow rate and mass loading \citep[see discussion in][]{MartinEtAl2013}. Consequently, comparisons with outflow rates derived from HVCs should be approached with caution.

Detections of both inflowing and outflowing gas in the halo of the Milky Way have lent support to the existence of a galactic fountain, where gas ejected from the disk by feedback processes eventually returns back to the disk together with new gas accreted from the CGM \citep{Fox_2019, Werk_2019, Marasco_2022}. The outflows seen in our simulated galaxies with CRs are also consistent with such a scenario, as we see both inflowing and outflowing gas that aids in the mixing of colder gas into the CGM.

\subsection{Caveats of the magnetic field structure}

We note that the initial conditions of the CGM are not self-consistently generated from cosmological initial conditions. Besides the simplified gas structure, the magnetic field strength and structure were chosen based on simplified assumptions. In our case, we bracketed the most likely CGM properties by a strong magnetic field model with a field strength of $\approx0.1\,\mu\mathrm{G}$ at a height of $10\,\mathrm{kpc}$ and a weak field with negligible field strength. In the former case, we found local  magnetically dominated regions and, therefore,  magnetically dominated dynamics as well. One artifact is the suppression of star formation at large galactocentric radii. The models MHD and CRMHD clearly show ring features that result from the strong initial fields. In addition, outflows are initially suppressed due to the stronger magnetic pressure in the CGM. The MHD-low and CRMHD-low models allow for a more self-consistent field amplification via the dynamo and, thus, a more naturally emerging dynamics. Whereas this is desirable in general, we also find caveats in terms of the dynamics. Since the growth of the field -- particularly at large altitudes above the plane -- takes much more time and exceeds the simulation time, the initial CR-driven feedback is unnaturally high. The mass-loading factors exceed 10 in the initial phase of the simulation, which allows  the halo to get filled up with dilute gas because of a non-existing resistive energy component in the CGM.

However, overall the extreme choices of the magnetic field do not overshadow the CR-driven dynamics. First of all, the main differences in the dynamics are driven by CRs, rather than the magnetic field. For the outflows and the resulting loading factors, we find agreement between both magnetic models. After an initial phase, both models converge to similar values in terms of the ability to launch an outflow from the disk. In the pure MHD case, this effect is absent. Furthermore, the effective magnetic field strength in the CGM is comparable after the first half of the simulation by a combination of magnetic dynamo and the transport of magnetized outflows from the disk. We conclude that the choice of the magnetic field is important for numerous details, but it is subdominant when comparing pure MHD models with CR counterparts.

\subsection{Caveats of CR transport and methods}\label{subsec: caveats}

The mechanisms of CR transport make up a complex, multiphysics problem to untangle. In our simulations, we use the simplified setup of one-moment CR transport in the diffusion-advection approximation, which requires some caveats. Although we account for the energy losses due to CR streaming \citep{wiener2013}, we do not explicitly model the interplay between the resonant waves driven by CRs and the scattering of CRs on these self-induced waves. Theoretical works have investigated the excitation of plasma instabilities and their impact on CR transport \citep{lemmerz2024} and found that the strength of CR driven winds might be greatly enhanced due to a stronger CR pressure gradient that results from CRs scattering off of self-induced instabilities \citep{Kulsrud1969,Shalaby_2021,Shalaby_2023}. The escape of CRs into the ISM is also affected, as the CR diffusion speed should be suppressed in the vicinity of the CR source, which would lead to an increased production of secondaries \citep{Shalaby_2021, Schroer_2022}.

There have already been efforts to implement a more realistic two-moment treatment of CR transport \citep{Jiang2018,thomas2019} that accounts for both CR streaming and CR diffusion. A two-moment approach was introduced to also self-consistently calculates the effective transport speed of CRs from the scattering of CRs off resonant Alfvén waves \citep{thomas2019,Thomas2022,Thomas2021}, finding that CR transport inside the galactic wind follows a distinctly non-steady state prescription \citep{Thomas2023}.

We also did not account for the energy-dependence of CR transport, which is being captured in novel approaches that follow the entire CR energy spectrum \citep{girichidis_I_2020}. CRs at different energies will have different transport speeds and cooling rates and \cite{Girichidis2024} found that the effective diffusive coefficient varies by two orders of magnitude in space and time, although the SFR and outflow rate are not strongly impacted.

As  mentioned in Section \ref{subsec:sim details}, we had mass return enabled in our simulations, implying that we did not allow for the depletion of the gas reservoirs in the galaxies over time. Instead, we achieved something more akin to a steady-state solution where global quantities such as the SFR (Figure \ref{fig:sfr}) and energy ratios stay relatively constant over the 2 Gyr of evolution. For the purpose of investigating the impact of CRs this is an advantageous setup, as CRs act on long timescales.

\section{Conclusions}\label{sec:conclusions}
In this work, we  investigate the dynamical impact of CRs on Milky Way-like galaxies using four simulations from the Rhea-suite. Two of the simulations (MHD and MHD-low) do not include the effect of CRs, while the other two (CRMHD and CRMHD-low) do. In the latter, the "-low" galaxies are initialized with a weaker magnetic field of $B_0=3\times10^{-9}$ G, compared to $B_0=3\times10^{-6}$ G for the remaining two. The simulations are performed using the moving-mesh code \textsc{Arepo} and include a non-equilibrium ISM model that accurately accounts for the relevant heating and cooling processes, a gravitational potential that matches the rotation curve of the Milky Way, and a star formation recipe designed so that the SFR is comparable to what we see in the Milky Way. CRs were injected in individual supernova explosions as 10\% of the explosion energy and  transported in the advection-diffusion approximation with an anisotropic diffusion coefficient along the magnetic field lines.  Streaming losses were emulated based on the gradient of the CR pressure. Our main conclusions can be summarized as follows:

\begin{enumerate}
    \item Overall, CRs help convert galactic fountain flows into outflows, which are not present in the pure MHD simulations. We find low-density and high-velocity outflows launched from the galactic center, which form a low-density cone into the CGM. In addition, we also observe outflows from the entire disk rising into the CGM, with outflow rates of around $\rm 10^5 \ M_\odot/Myr$. The mass-loading factors of the outflow converge for the strong and weak magnetic field case after approximately half of the simulation time with values around 0.2.
    \item CRs lead to strongly magnetized ($\beta=0.1-1$) and colder ($T\sim10^4\,\mathrm{K}$) CGM, compared to low magnetized ($\beta\sim10^3$) and hot ($T=10^6\,\mathrm{K}$) CGM in their MHD counterparts. We note second-order differences between the low and high magnetic field case. In the low-$B$ setup, CRs and magnetized gas are transported faster into the CGM at early times due to the lack of resistance in the CGM. 
    \item The distribution of CRs is dominated by non-adiabatic processes. An adiabatic scaling of the CR energy density $e_\mathrm{cr}\propto\rho^{4/3}$ is observed only at the lowest densities. For a significant density range above $\rho\gtrsim10^{-24}\,\mathrm{g\,cm}^{-3}$, we find a flat distribution, suggesting that fast transport dominates the spatial distribution. This is in line with the analyzed CR gradient lengths of $L_{\mathrm{cr}} = e_\mathrm{cr} / \boldsymbol{\nabla} e_\mathrm{cr} \gtrsim 100\,\mathrm{pc}$, which show that the CRs quickly diffuse into a smooth distribution with weak gradients and, consequently, weak CR pressure forces, which are unable to stop the gravitational collapse of dense clouds.
    \item The energy transported in the outflows of the two CR simulations predominantly takes the form of CR energy. We find that 18\% and 58\% of the injected CR energy escapes in the outflow in CRMHD and CRMHD-low, respectively. The stronger magnetic field in the CRMHD simulation results in more CRs staying inside the galaxy, where they can experience energy losses through interactions with the ISM.
\end{enumerate}

In summary, our findings highlight the central role of CRs in shaping the evolution of Milky Way-like galaxies, especially in driving outflows, which are absent in the MHD counterparts, and influencing the vertical structure. While magnetic fields also contribute to these processes, their impact is secondary to that of CRs.

\begin{acknowledgements}
\karin{We thank the anonymous referee for a careful reading of the manuscript and constructive comments that helped improving the paper.}
The team in Heidelberg  acknowledges financial support from the European Research Council via the ERC Synergy Grant ``ECOGAL'' (project ID 855130),  from the German Excellence Strategy via the Heidelberg Cluster of Excellence (EXC 2181 - 390900948) ``STRUCTURES'', and from the German Ministry for Economic Affairs and Climate Action in project ``MAINN'' (funding ID 50002206).00
The authors gratefully acknowledge the scientific support and HPC resources provided by the Erlangen National High Performance Computing Center (NHR@FAU) of the Friedrich-Alexander-Universität Erlangen-Nürnberg (FAU) under the NHR project a104bc. NHR funding is provided by federal and Bavarian state authorities. NHR@FAU hardware is partially funded by the German Research Foundation (DFG) – 440719683.
They also thank for computing resources provided by the Ministry of Science, Research and the Arts (MWK) of the State of Baden-W\"{u}rttemberg through bwHPC and the German Science Foundation (DFG) through grants INST 35/1134-1 FUGG and 35/1597-1 FUGG, and for data storage at SDS@hd funded through grants INST 35/1314-1 FUGG and INST 35/1503-1 FUGG.
KK, PG, JG, NB, and RSK acknowledge financial support from the European Research Council (ERC) via the ERC Synergy Grant ``ECOGAL'' (grant 855130), from the German Excellence Strategy via the Heidelberg Cluster of Excellence (EXC 2181 - 390900948) ``STRUCTURES'', and from the German Ministry for Economic Affairs and Climate Action in project ``MAINN'' (funding ID 50OO2206). 
KK and JG are fellows of the International Max Planck Research School for Astronomy and Cosmic Physics at the University of Heidelberg (IMPRS-HD).
NB acknowledges support from the ANR BRIDGES grant (ANR-23-CE31-0005).
RSK also thanks the Harvard-Smithsonian Center for Astrophysics and the Radcliffe Institute for Advanced Studies for their hospitality during his sabbatical, and the 2024/25 Class of Radcliffe Fellows for highly interesting and stimulating discussions. CP acknowledges support by the European Research Council under ERC-AdG grant PICOGAL-101019746.

\end{acknowledgements}

\bibliographystyle{aa} 
\bibliography{references}

%\clearpage
\onecolumn
\begin{appendix}
\section{Resolution}\label{app:resolution}
\begin{figure*}[h!]
\centering
    \includegraphics[width=\textwidth]{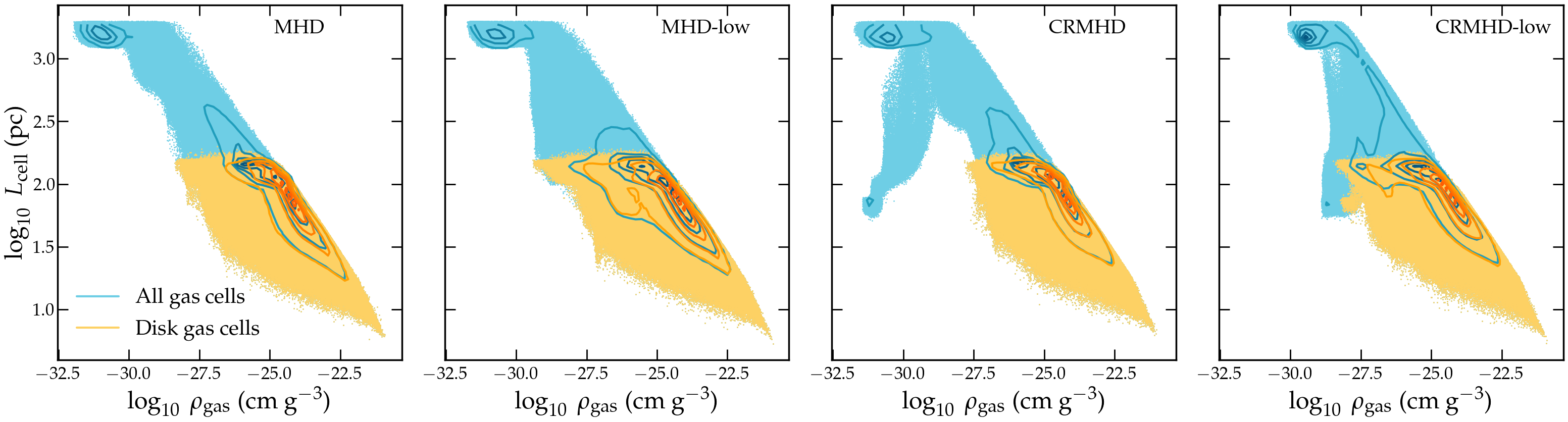}
    \caption{Cell length-density distribution of our simulations at $t=1.5$ Gyr. In blue we plot all the gas cells, while in yellow only those belonging to the disk ($r<20$ kpc, $h< 1$ kpc). The contours enclose 10\%, 30\%, 50\%, 70\%, and 90\% of the cells in that region.}
    \label{fig:cell-length-density}
\end{figure*}

To confirm the resolution of our simulations, we plot in Figure~\ref{fig:cell-length-density} the cell size versus the gas density for all our models at a time of $t = 1.5$ Gyr. The blue and yellow regions denote cells in the entire box and cells in the disk, respectively. The contours enclose 10\%, 30\%, 50\%, 70\%, and 90\% of the cells in that region. The additional volume refinement in the disk, which limits the maximum cell volume to (100 pc)$^3$, is evident, resulting in reduced cell lengths in this area. \karin{In the CRMHD-simulation there is a collection of cells outside the disk with low density and high resolution, which are cells that were dislodged from the galactic disk and have not had time to de-refine yet.}

\FloatBarrier
\clearpage

\section{Magnetic field scaling}\label{app:Bfield_scaling}
\begin{figure*}[h!]
    \centering
    \includegraphics[width=\textwidth]{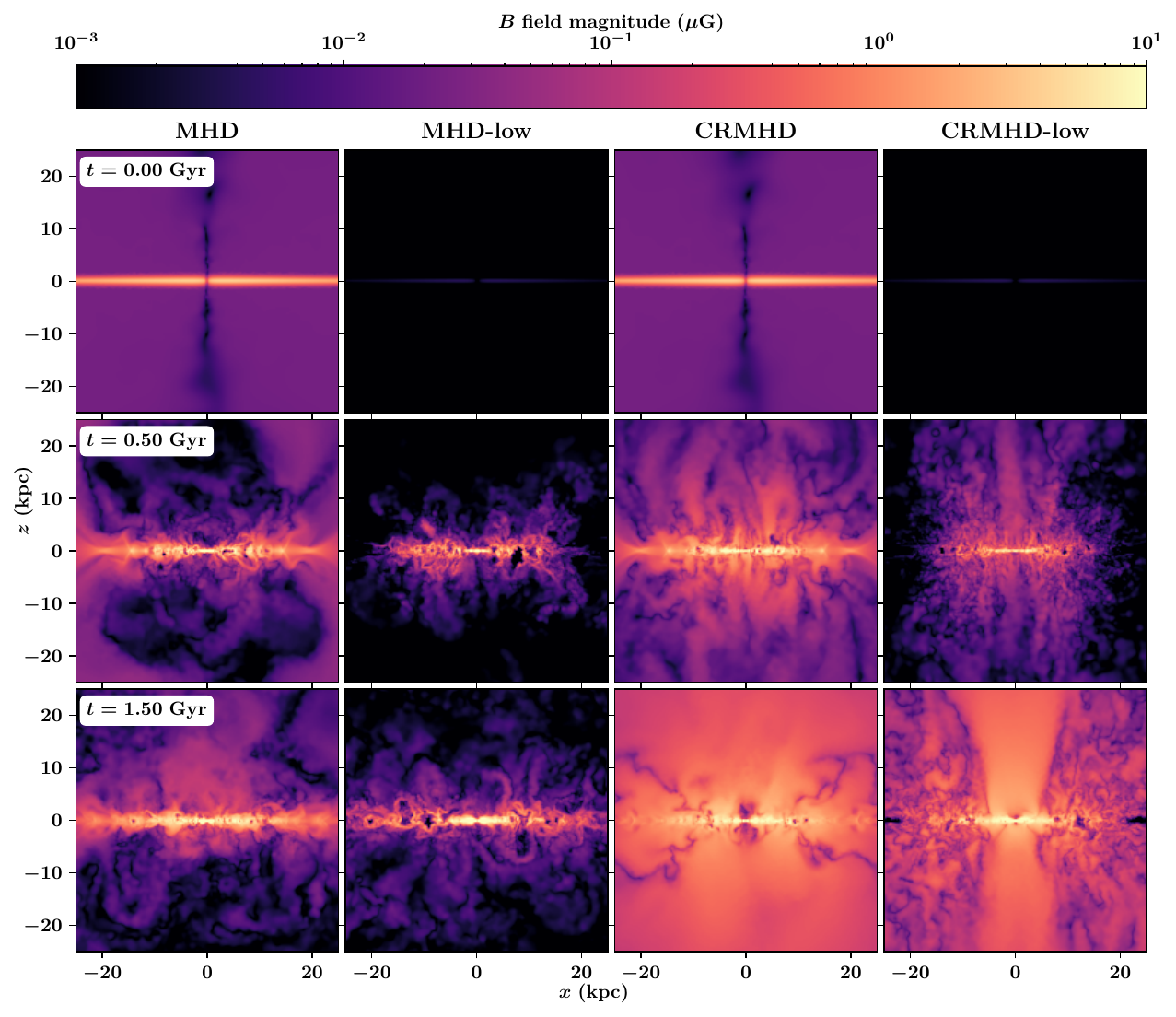}
    \caption{Edge-on view of the magnetic field strength for all models (from left to right) for different times (top to bottom). The initially strong magnetic field in models MHD and CRMHD only leave a weak imprint after 1\,Gyr.}
    \label{fig:Bfield-CGM-time-evol}
\end{figure*}

We use two different initial magnetic field strengths in the simulation setup. Here we address the temporal evolution of the magnetic field in the CGM as well as the resulting outflow properties over time. In Figure~\ref{fig:Bfield-CGM-time-evol} we show the edge-on view of the magnetic field for all four simulations. The different initially strong field in models MHD and CRMHD are only clearly visible at early times. After approximately 1\,Gyr of evolution, the dynamics in the CGM is dominated by CR-driven outflows from the disk. Therefore, the magnetic field strength shows main differences between the CR models and the MHD only models, evident from the last row of Figure~\ref{fig:Bfield-CGM-time-evol} which depicts the galaxies after 1.50 Gyr of evolution. CR-driven outflows result in stronger fields of order $\mu\mathrm{G}$ throughout the depicted CGM.

\FloatBarrier
\clearpage

\section{Outflows at $h=10$ kpc}\label{app:outflows_10kpc}
\begin{figure*}[h!]
    \includegraphics[width=\textwidth]{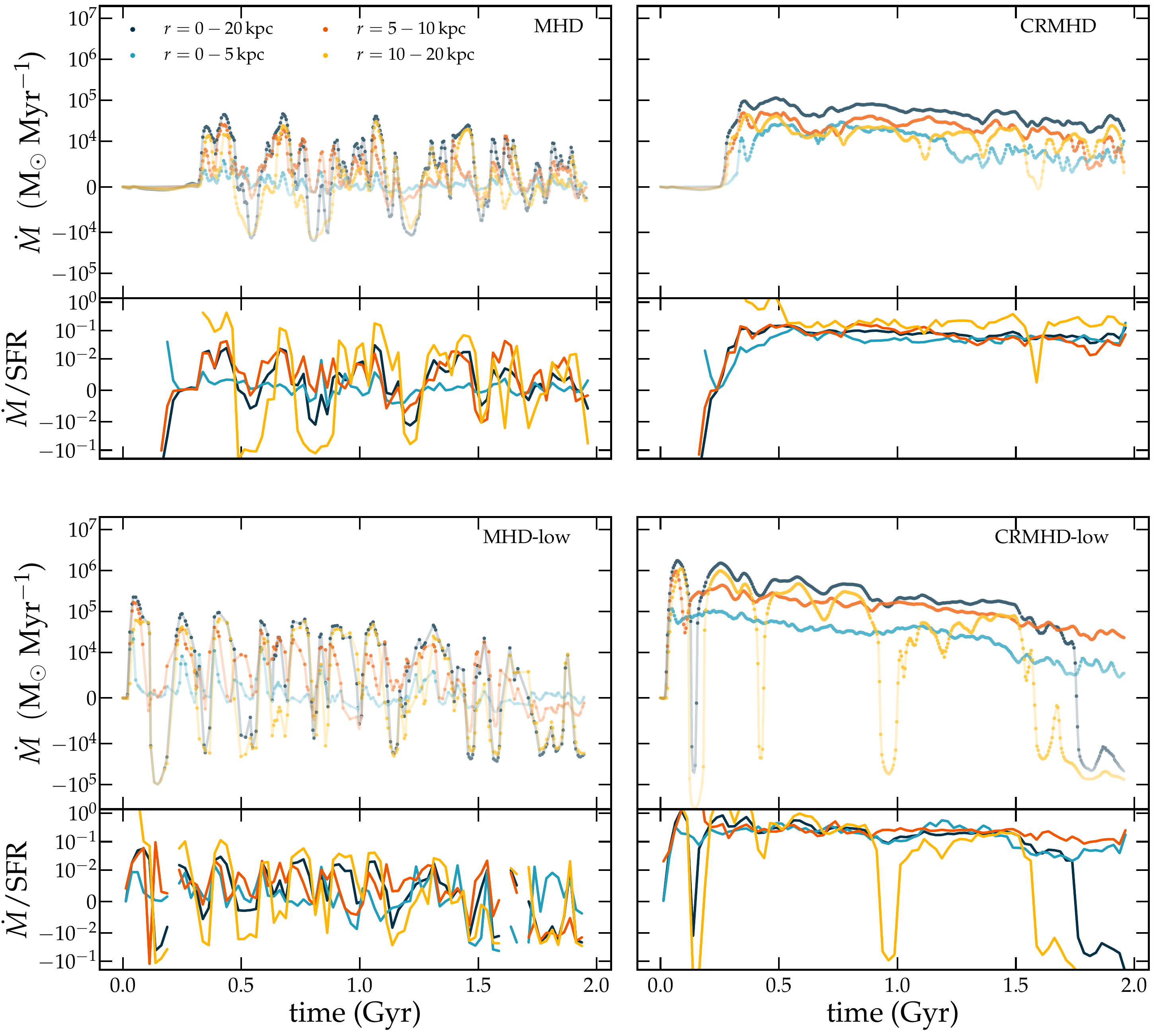}
    \caption{Mass outflows at $|z|=10$ kpc for our four simulated galaxies. The upper panels show the mass outflow rate as a function of time, with different colors signifying different radial bins. The lower panels show the mass outflow rate scaled by the SFR, also known as the mass-loading factor. }
    \label{fig:outflows10kpc}
\end{figure*}

Figure \ref{fig:outflows10kpc} shows mass outflow rates (upper panels) and mass-loading factors (lower panels) at a height of 10 kpc above the disk in different radial bins, for all our four simulations. Outflows are calculated the same way as for Figure \ref{fig:outflows}. The MHD simulations exhibit no coherent outflows, as was the case at $h=5$ kpc, instead having mass flux rates fluctuating between infalling and outflowing. Outflows in the CRMHD-galaxy are generally weaker at this larger height than at $h=5$ kpc but still exhibit a net outflow in all radial bins. Notably, in the CRMHD-low simulation, there are consistent outflows after 1.5 Gyr of evolution in the 5-10 radial bin, whereas these outflows were seen to disappear at a height of 5 kpc. The CR-driven outflows generate large-scale perturbations up to heights of around $h\approx5$ kpc, which leads to infalling gas after $t=1.5$ Gyr. However, gas that reaches to 10 kpc continues moving outward, producing weaker but coherent outflows at this height.

\end{appendix}

\end{document}